\documentclass[aps,pre,twocolumn,superscriptaddress,showpacs,showkeys,longbibliography]{revtex4-1}
\usepackage[english]{babel}
\usepackage{wrapfig,chngcntr}
\usepackage{amsmath, amsfonts, amssymb, graphicx, float}
\usepackage{amsbsy,amsthm}
\usepackage{algorithm,algorithmic}
\usepackage{mathrsfs,multirow}
\usepackage{bm,comment}
\usepackage{verbatim}
\usepackage[caption=false]{subfig}
\usepackage{algorithm}
\usepackage{url,color}
\usepackage{latexsym}

\newtheorem{property}{Property}
\usepackage[final]{changes}

\pdfoutput=1 

\begin{document}

\title{Weight Thresholding on Complex Networks}

\author{Xiaoran Yan}\email{xiaoran.a.yan@gmail.com}
\affiliation{Indiana University Network Science Institute (IUNI), 1001 E SR 45/46 Bypass, Bloomington, IN 47408, USA}
\author{Lucas G. S. Jeub}
\affiliation{School of Informatics, Computing and Engineering, Indiana University, 700 N Woodlawn Ave, Bloomington, IN 47408, USA}
\author{Alessandro Flammini}
\affiliation{Indiana University Network Science Institute (IUNI), 1001 E SR 45/46 Bypass, Bloomington, IN 47408, USA}
\affiliation{School of Informatics, Computing and Engineering, Indiana University, 700 N Woodlawn Ave, Bloomington, IN 47408, USA}

\author{Filippo Radicchi}
\affiliation{School of Informatics, Computing and Engineering, Indiana University, 700 N Woodlawn Ave, Bloomington, IN 47408, USA}

\author{Santo Fortunato}
\affiliation{Indiana University Network Science Institute (IUNI), 1001 E SR 45/46 Bypass, Bloomington, IN 47408, USA}
\affiliation{School of Informatics, Computing and Engineering, Indiana University, 700 N Woodlawn Ave, Bloomington, IN 47408, USA}

\begin{abstract}
Weight thresholding is a simple technique that aims at reducing the number of edges in weighted networks that are otherwise too dense for the application of standard graph-theoretical methods. We show that the group structure of real weighted networks is very robust under weight thresholding, as it is maintained even when most of the edges are removed. This appears to be related to the correlation between topology and weight that characterizes real networks. On the other hand, the behavior of other properties is generally system dependent. 

\end{abstract}
\maketitle
%
%
\section{Introduction}

Many real networks have weighted edges~\cite{barrat04}, representing the intensity of the interaction between pairs of vertices. Also, some weighted networks, e.g., financial~\cite{namaki2011network} and brain networks~\cite{bullmore_complex_2009}, have a high density of edges. Analyzing very dense graphs with tools of network science is often impossible, unless some pre-processing technique is applied to reduce the number of connections. Different recipes of edge pruning, or {\it graph sparsification}, have been proposed in recent years~\cite{tumminello05,serrano_extracting_2009,radicchi11,spielman_resistance_2011,Dianati2016,coscia2017backboning,2018arXivKobayashi}.

In practical applications such as the pre-processing of data about brain, financial and biological networks \cite{lynall_functional_2010,allesina_secondary_2006,namaki2011network}, weight thresholding is the most popular approach to sparsification. It consists in removing all edges with weight below a given threshold. Ideally, one would like to eliminate as many edges as possible without drastically altering key features of the original system. 
A recent study of functional brain networks investigates how the graph changes as a function of the threshold value~\cite{garrison_stability_2015}, finding that conventional network properties are usually disrupted early on by the pruning procedure. Further, many standard measures do not behave smoothly under progressive edge removal, and hence are not reliable measures to assess the effective change to the system structure induced by the removal of edges.  

By analyzing several synthetic and real weighted networks, we show that, while local and global network features are often quickly lost under weight thresholding, the procedure does not alter the mesoscopic organization of the network~\cite{fortunato2010community}: groups (e.g., communities) survive even when most of the edges are removed. 

\added{In addition, we introduce a measure, the {\it minimum absolute spectral similarity} (MASS), that estimates the variation of spectral properties of the graph when edges are removed. Spectral properties are theoretically related to group structures in general \cite{luxburg_tutorial_2007,sarkar_eigenvector_2015,Iriarte2016}. The MASS is stable under weight thresholding for many real networks, making it a potential alternative to expensive community detection algorithms for testing the robustness of communities under thresholding.}


\section{Methods}

Let us consider a weighted and undirected graph $G$ composed of $N$ vertices and $M$ edges. Edges have positive weights, and the graph topology is described by the symmetric weight matrix $W$, where the generic element $W_{uv} = W_{vu} > 0$ if there is a weighted edge between nodes $u$ and $v$, while $W_{uv} = W_{vu} = 0$, otherwise. Weight thresholding removes all edges with weight lower than a threshold value. This means that the resulting graph $\tilde{G}$ has a thresholded weight matrix $\tilde{W}$, whose generic element $\tilde{W}_{uv} = \tilde{W}_{vu} = W_{uv}$ if $W_{uv} \geq \theta$, and $\tilde{W}_{uv} = \tilde{W}_{vu} = 0$, otherwise. The thresholded graph $\tilde{G}$ is therefore a subgraph of $G$ with the same number of nodes.

We first examined synthetic networks generated by the Lancichinetti, Fortunato, and Radicchi (LFR) benchmark \cite{2008LFR}. We then extend the analysis to several real networks from 
different domains: structural brain networks \cite{betzel2013multi}, the world trade network \cite{de2011trade}, the airline network \cite{openFlights}, and the co-authorship network of faculty of Indiana University. We treat all networks as undirected, weighted graphs with positive edge weights. We show the variation of several graph properties as a function of the fraction of removed edges. The properties we have chosen include all network-level measures from Ref.~\cite{garrison_stability_2015} as well as mesoscopic structure measures:

\begin{enumerate}
 \item {\it Characteristic path length (CPL)}, the average length of all shortest paths connecting pairs of vertices of the network. Once the network becomes disconnected, CPL is not defined, and we set its value to $0$.

 \item {\it Global efficiency}, the average inverse distance between all pairs of vertices of the network~\cite{Latora2001Efficient}. Global efficiency remains well defined after network disconnection, as disconnected vertex pairs simply have a inverse distance of $0$. 
 
 \item {\it Transitivity}, or global clustering coefficient, is the ratio of triangles to triplets in the network, where a triplet is a motif consisting of one vertex and two links incident to the vertex.
 
 \item {\it Community structure}, which We detect with two methods. The first is based on modularity maximization~\cite{newman_modularity_2006} via the Louvain algorithm \cite{Blondel2008Louvain, mucha_community_2010}. The second method uses $k$-means spectral clustering \cite{luxburg_tutorial_2007} with a constant $k$ equal to the one
 found with the Louvain algorithm. 
 
 \item {\it Core-periphery structure}. We detect the bipartite partition of core and periphery vertices using the method introduced for the weighted coreness measure \cite{Rubinov2015}. 

 \item {\it DeltaCon}, a metric indicating the similarity between the original graph and the thresholded one based on graph diffusion properties ~\cite{Koutra_DeltaCon_2016}. 
\end{enumerate}

For all three partitioning algorithms, i.e., Louvain, $k$-means and coreness, we measure the similarity between the partitions of the sparsified graph $\tilde{G}$ and those of $G$ using the {\it adjusted mutual information} (AMI)~\cite{vinh2010information}. To overcome the randomness of the partitioning algorithms, we sample 100 different partition outcomes from the original graph, 10 from the sparsified graph, and use the maximum AMI between any pair. These numbers are picked so that the same algorithm returns consistent results (AMI very close to 1) on independent runs on the original graph.

For global efficiency we take the ratio between the value of the measure on the sparsified graph and the corresponding value on the initial graph. CPL grows with edge removal, and we therefore normalize it by its largest value before disconnection. Since DeltaCon scales directly with weights, we normalize each matrix entry such that it reaches $0$ on empty graphs. This way all our measures are confined in the interval $[0,1]$, and their trends can be compared (transitivity naturally varies in this range). We also plot the relative size of the largest connected component, to keep track of splits of the network during edge removal. More details of how these graph properties are calculated are given in Appendix A.

We also add another measure, capturing the variation of spectral properties of the graph. To define this measure we recall that the Laplacian of $G$ is defined as $L = D-W$, where $D$ is the diagonal matrix of the weighted degrees (strengths), with entries $D_u = \sum_v W_{uv}$. Similarly, for the sparsified graph 
$\tilde{G}$, $\tilde{L} = \tilde{D}-\tilde{W}$.
The Laplacian has the spectral decomposition:
\[
 L = D - W = V\Lambda V^{-1}\;,
\]

where the columns of the matrix $V$ are the eigenvectors $v_1, v_2,  \dots, v_N$ of the Laplacian, and the entries in the diagonal matrix $\Lambda$ are the corresponding eigenvalues $0=\lambda_1\le\lambda_2\le...\le\lambda_N$. 

The difference between the sparsified Laplacian $\tilde{L}$ and the original $L$, can be quantified by the \emph{minimum relative spectral similarity} (MRSS)~\cite{spielman_spectral_2008},
\begin{align}
\label{eq:MRSS}
  \sigma_{min}^R = \min_{\forall x}\frac{x^T\tilde{L}x}{x^TLx}\;,
\end{align}
where $x^T\tilde{L}x$ is the Laplacian quadratic form and $x$ any $N$-dimensional real vector. The MRSS is a direct adaptation of relative spectral bounds. Intuitively, the input vector $x$ determines the ``direction'' along which we measure the change of the graph, and by taking the minimum we consider the worst case scenario. However, the value of MRSS drops to zero as soon as $\tilde{G}$ becomes disconnected. Because of this mathematical degeneracy, it is also numerically unstable for many optimization algorithms.

To overcome this issue, we instead propose the {\it absolute spectral similarity} with respect to the input vector $x$, defined as 
\begin{align}
\label{eq:ass}
\sigma(x) = 1-\frac{x^T\varDelta Lx}{x^T \lambda_N x}\;,
\end{align}
where $\lambda_N$ is the largest eigenvalue of the original graph Laplacian, and $\varDelta L = \varDelta D - \varDelta W$ is the graph Laplacian of the {\it difference graph} $\varDelta G$, whose vertices are the same as in $G$, while the edges are the ones removed by the chosen sparsification procedure. Without loss of generality, we consider only unit length input vectors, $|x| = 1$. In Appendix B, we prove that weight thresholding optimizes the expected value of $\sigma(x)$, if we assume the entries of the vector $x$ are independent identically distributed random variables.


Since the input vector is variable, we again consider the worst case scenario and use the {\it minimum absolute spectral similarity} (MASS),
\begin{eqnarray}
\label{eq:minSS}
 \sigma_{min} &=& \min_{|x| = 1}\left(1 - \dfrac{x^T\varDelta Lx}{\lambda_N}\right) = \dfrac{\lambda_N-\lambda_N^{\varDelta}}{\lambda_N} \;,
\end{eqnarray}
where $\lambda_N^{\varDelta}$ is the largest eigenvalue of the difference Laplacian $\varDelta L$. 

MASS is consistent with our intuition that disconnecting a few peripheral vertices, while a bigger change than removing redundant edges, should have a small impact on the organization of the system. Many local and global network features become ill-defined as soon as the network disconnects, and hence they are not reliable measures to assess the effect of thresholding. Mesoscopic properties, like communities, on the other hand, remain meaningful even after the network becomes disconnected, and a reliable measure should be robust in such situations.

A major advantage of MASS is its numerical stability and computational efficiency (see Appendix C). Defined as the ratio of the largest eigenvalues of two graph Laplacians, it can be computed using standard numerical libraries\cite{sorensen1992implicit,1998_arpack,Stewart_2001}. \added{(Readers can find our Matlab implementation at https://github.com/IU-AMBITION/MASS.)} In spite of its simplicity, MASS satisfies a series of theoretical properties (see Appendix E).

\begin{figure*}
\centering
 \includegraphics[width=0.95\linewidth]{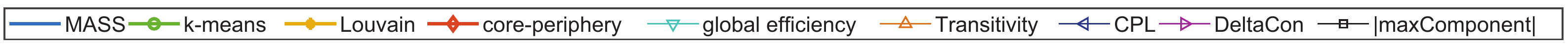}\\
{\includegraphics[width=0.48\linewidth]{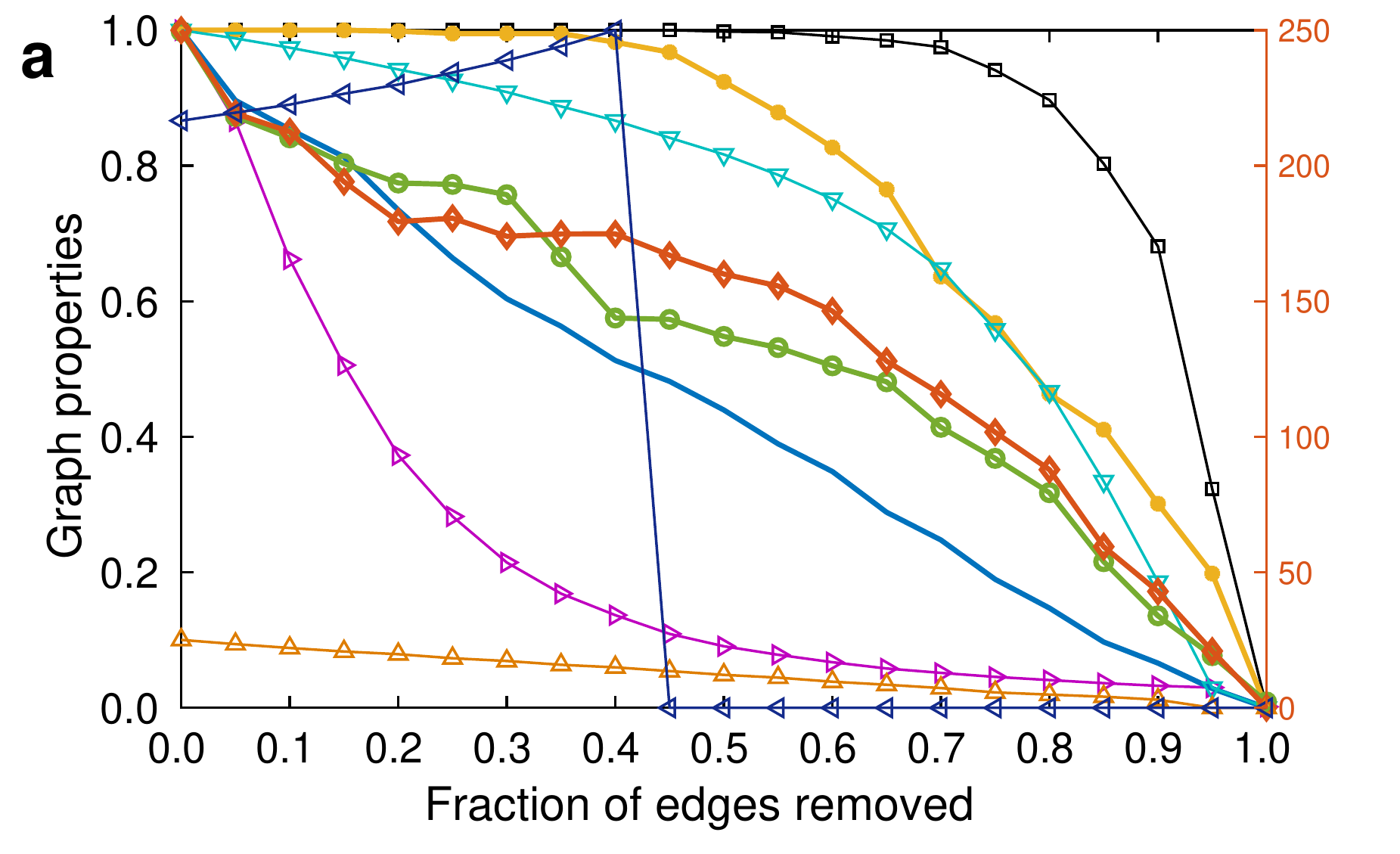}}
{\includegraphics[width=0.48\linewidth]{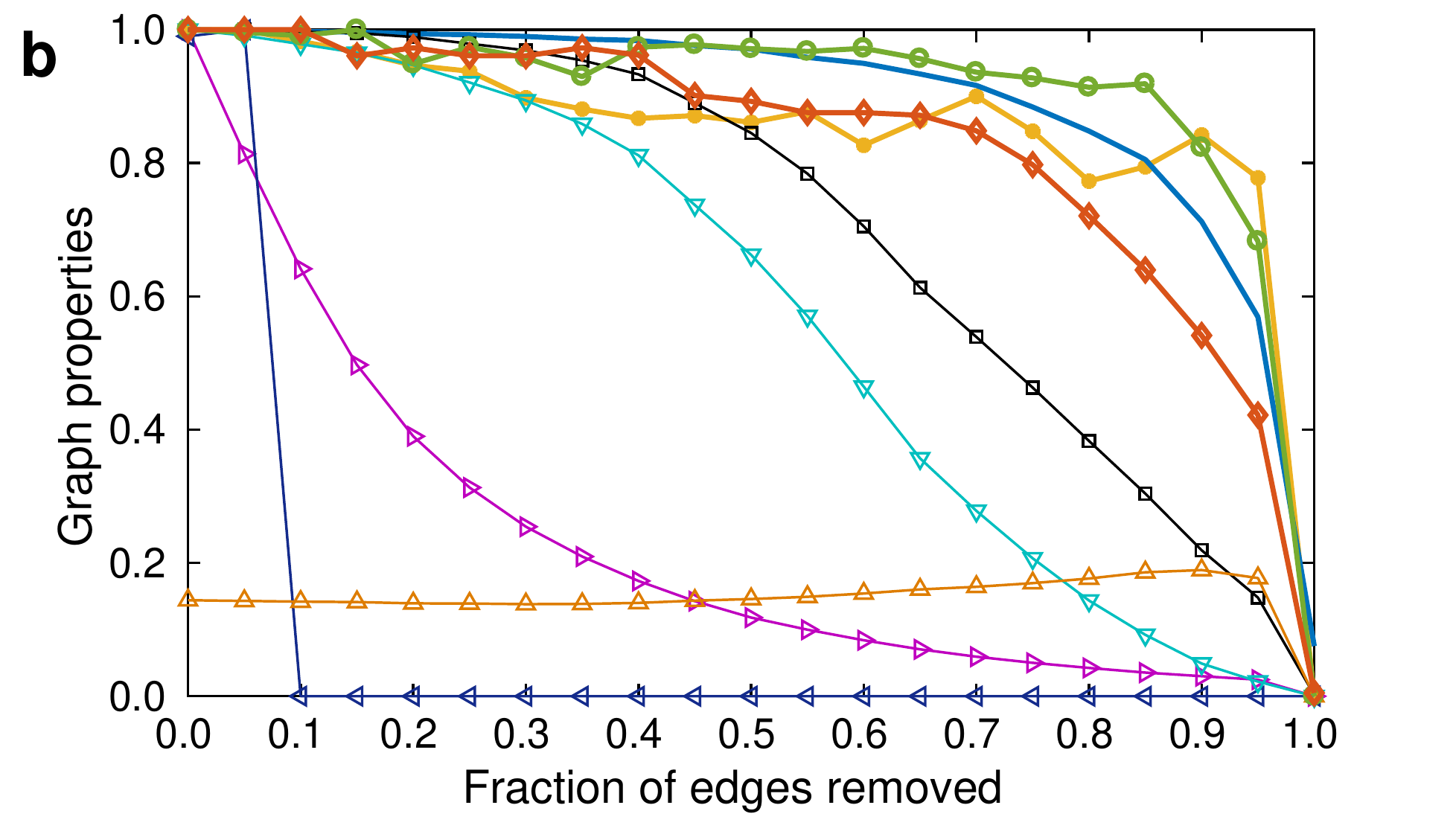}}
\caption{Variation of MASS and other variables under weight thresholding, for synthetic binary networks generated by the LFR benchmark (a) and a weighted counterpart with noisy correlation between edge weight and the degrees of its endpoints (b). The "$|$maxComponent$|$" curve represents the fraction of vertices remaining in the largest connected component.}
\label{fig:LFR}
\end{figure*}

\section{Results}

To illustrate the changes of the above graph properties under weight thresholding, we first compare a synthetic weighted network with its binary counterpart, see Fig.~\ref{fig:LFR}. The binary network with 1000 vertices and 8 planted communities is generated using the LFR benchmark \cite{2008LFR} (the parameters of the LFR benchmarks used here are listed in Appendix A), which produces realistic networks by capturing power-law distributions of both degree and community size. Since all edges have the same weight, the thresholding is done by removing a fraction of randomly selected edges (In Fig.~\ref{fig:LFR}a, all curves are averaged over 10 random realizations). While the network remains well connected even after we remove $50\%$ of edges, the AMI scores relative to the mesoscopic group structures drop already when a small fraction of edges is removed. The MASS curve decays even faster, following a diagonal line, which suggests that it captures changes in spectral properties that are not directly related to the mesoscopic structure of the network.

In many real weighted networks, there is a power law relation between the degree of a vertex and its strength (i.e., the total weight carried by the edges adjacent to the vertex) \cite{2009LFweighted,barrat04}. We first set the weight of an edge to be proportional to the product of the degrees of its endpoints, changing the binary network into a weighted one. To account for the presence of noise we added a uniform error in the range of $[-w, w]$ on top of each weighted edge, with $w$ being the corresponding edge weight. The effect of thresholding on these networks is shown in Fig.~\ref{fig:LFR}b. As a result, the network's group structures are now robust under weight thresholding, despite the fact that a substantial fraction of vertices become disconnected at an early stage. In this realistic weighted network, the MASS curve now has a quite similar trend as the curves describing the variation of group structure. On the other hand the CPL becomes ill-defined as soon as the network is disconnected. DeltaCon also drops much earlier compared with other measures, as diffusion is heavily affected by network disconnection.

Let us now discuss the analysis of the real networks. We list their basic properties in Table~\ref{tab:datasets}. All four networks have positive weight-degree correlation coefficient, as it often happens in real weighted networks~\cite{barrat04}. Another observation is that they also have high values of the weighted coreness measure~\cite{Rubinov2015} (except for structural brain networks), which means weakly connected peripheral vertices will quickly become disconnected.

\begin{table*}
\caption{Networks studied in this paper and their properties}
\begin{center}
\setlength{\tabcolsep}{.25em}
\bgroup\def\arraystretch{1.5}
	\begin{tabular}{lccccccccc}
	\hline\hline
	\emph{Name} & \#\emph{samples}  & \#\emph{vertices}  & \#\emph{edges} & \#\emph{k (Louvain)} &\#\emph{Weight-degree correlation} &\#\emph{Coreness} \\
	\hline
Structural brain networks  &40		&234  		        &4046 (mean)	        &9.9 (mean)          &0.2603 (mean)     &0.3009 (mean)\\
	World trade network  	   &1   	&250  		        &18389	        &3                  &0.9539     &0.8373\\
	Airline network   	       &1   	&3253 		        &18997		    &20                 &0.5848     &0.6776\\ 
	Co-authorship network      &1   	&2855 		        &75058		    &7                  &0.6247     &0.8223\\
	\hline\hline
	\end{tabular}\egroup
\end{center}
\label{tab:datasets}
\end{table*}

The structural brain network is built from diffusion weighted imaging MRI scans of 40 experiment participants \cite{betzel2013multi}. Each personal network has 234 vertices representing brain regions. On average, they have 7862 weighted edges representing the fiber density connecting couples of regions. Here we threshold each network individually and considered the population average of each graph property for our analysis.

\begin{figure*}
\centering
 \includegraphics[width=0.95\linewidth]{legend2-eps-converted-to}\\
{\includegraphics[width=0.49\linewidth]{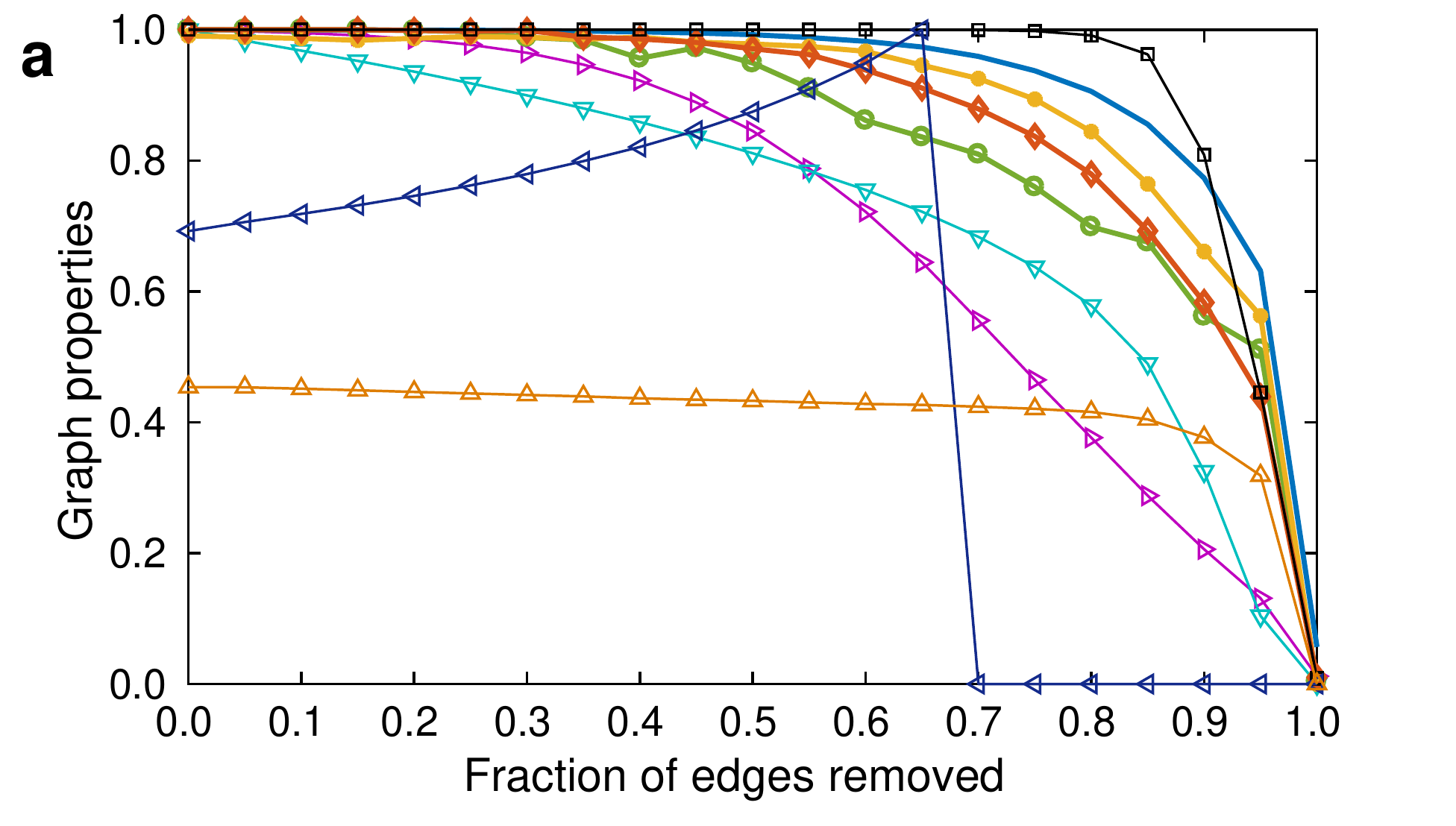}}\hfill
{\includegraphics[width=0.49\linewidth]{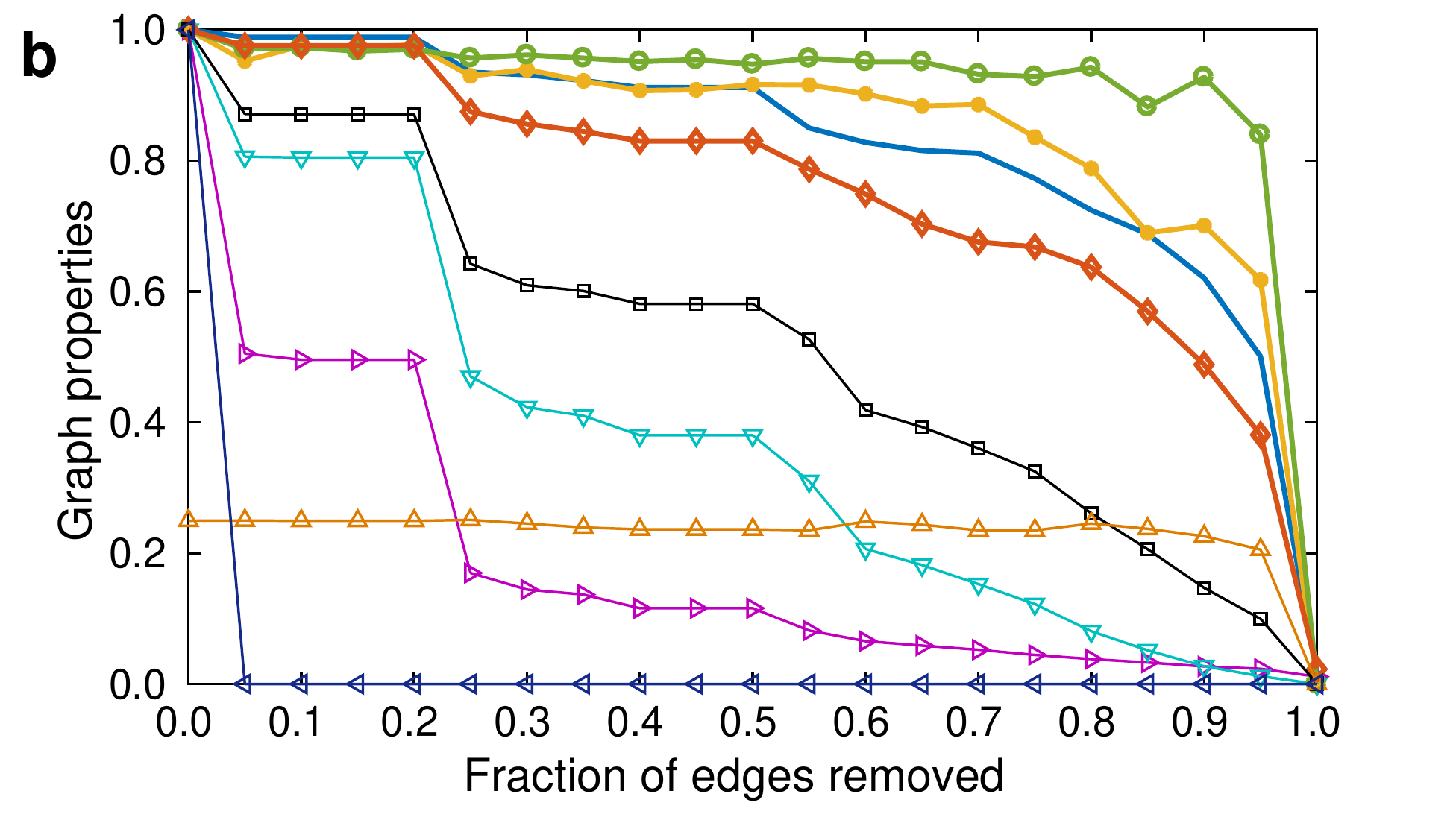}}
{\includegraphics[width=0.49\textwidth]{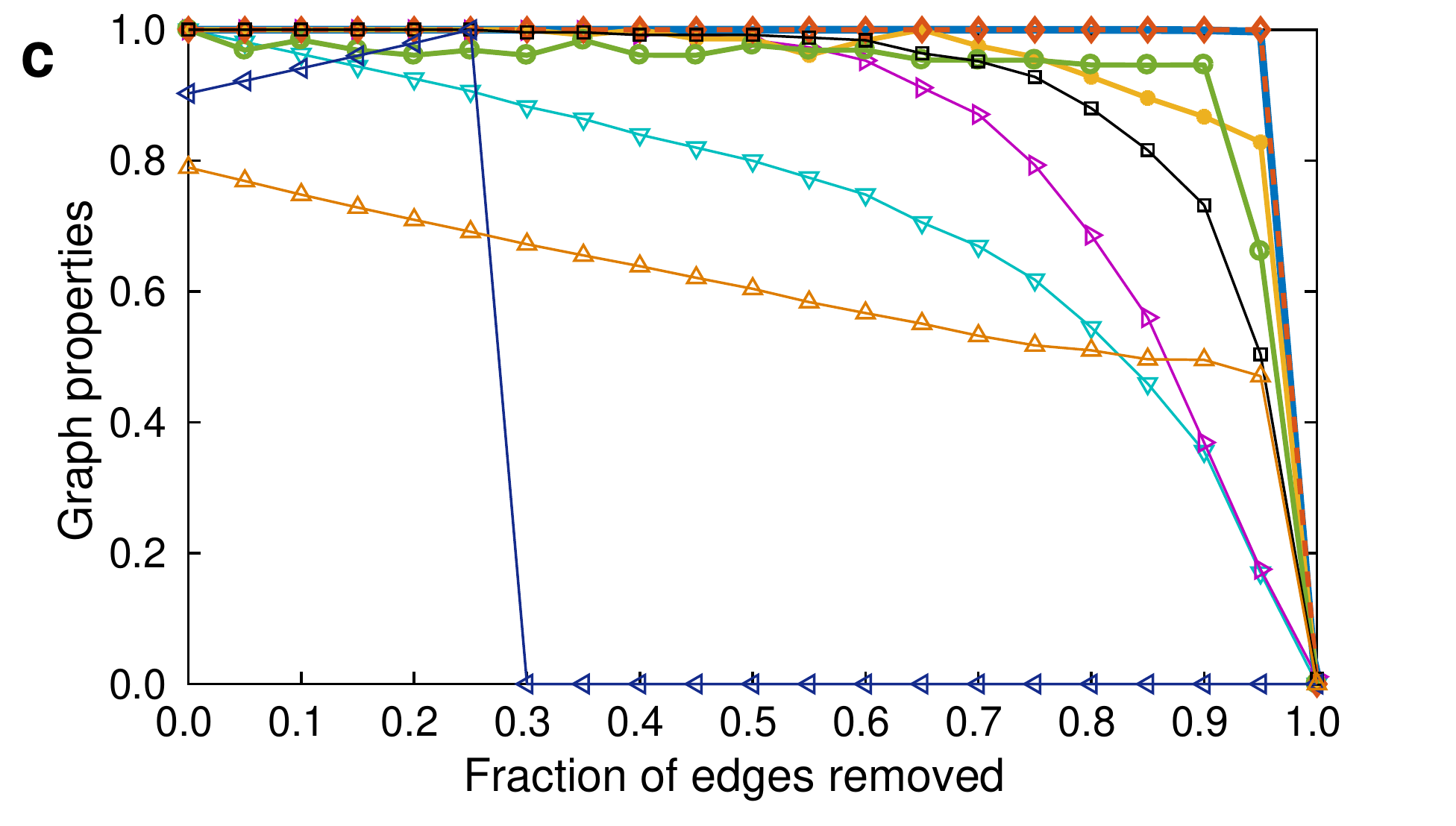}}\hfill {\includegraphics[width=0.49\textwidth]{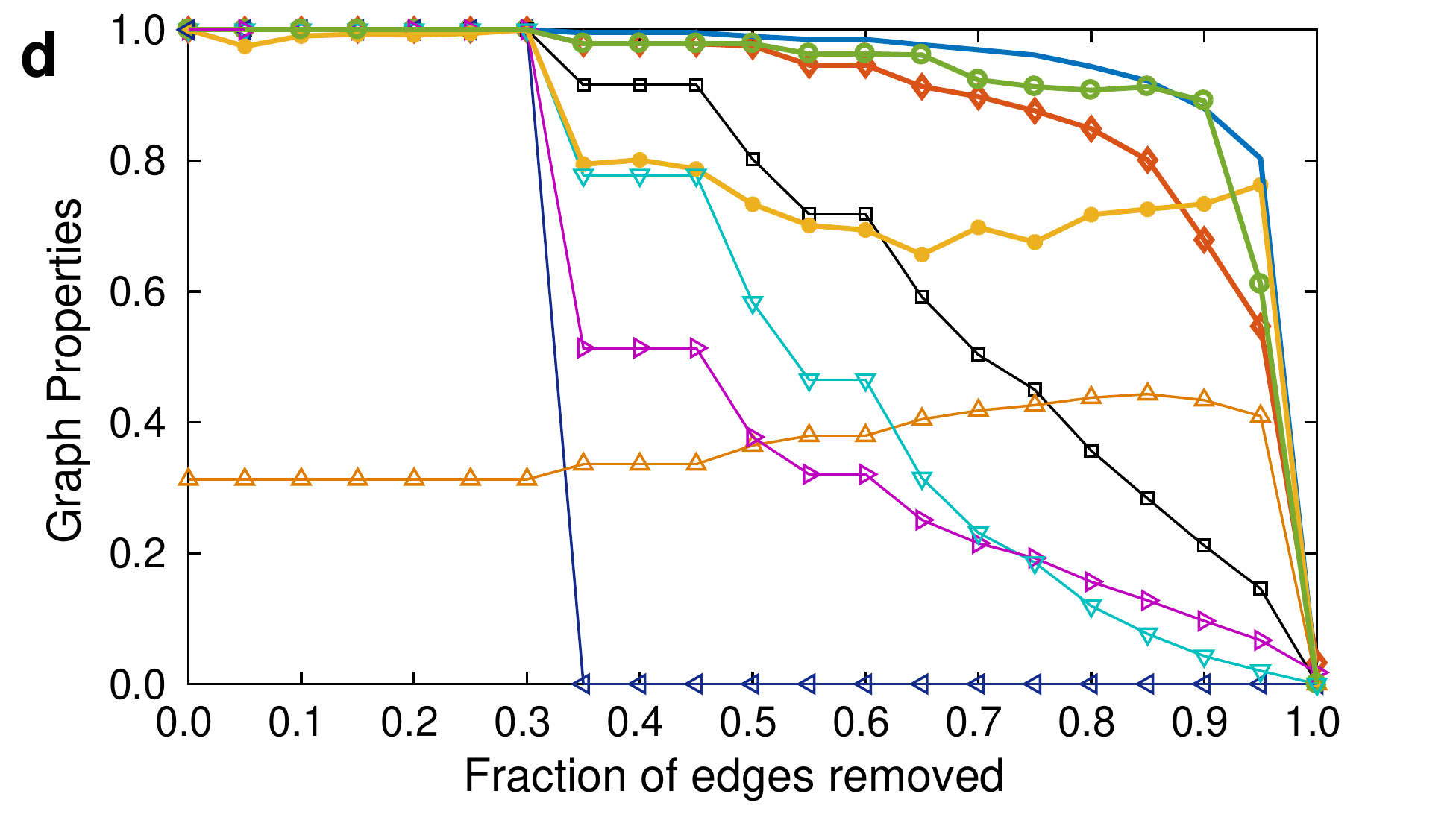}}

\caption{Variation of MASS and other variables under weight thresholding, for structural brain networks (a), the global airline network (b), the world trade network (c) and the co-authorship network (d).The "$|$maxComponent$|$" curve represents of the fraction of vertices remaining in the largest connected component.}
\label{fig:brain}
\end{figure*}

Fig.~\ref{fig:brain}a shows that the network is robust when the light edges are removed: it takes the removal of a substantial fraction of edges to get appreciable changes in all measures. In particular, the MASS is basically unaffected until more than half of the edges are deleted and it follows qualitatively the trend of the similarity (AMI) of the mesoscopic group partitions. 

The airline network is constructed from public data on flights between major airports around the world, with the edge weight representing the number of flights as well as the capacity of the plane operating each flight \cite{openFlights}. Small airports have only weak connections to the rest of the system, due to the limited traffic they handle. On the other hand, major hubs have some of the strongest connections, leading to a strong core-periphery structure with $0.5848$ weight-degree correlation. The system gets quickly disconnected under weight thresholding, and we see big drops in other graph properties early on, including CPL, DeltaCon and global efficiency, as shown in Fig.~\ref{fig:brain}b. Community and core-periphery structure, on the other hand, remain fairly stable against edge removal, and the MASS again follows a similar trend as the group partition similarity (AMI) curves. 

The world trade network is constructed from economic trading data between 250 countries. We aggregate the directed edges into 18389 undirected edges representing bidirectional trade volumes \cite{de2011trade}. This network is an extreme example of strong core-periphery structure with close to $1$ weight-degree correlation ($0.9539$) and coreness measure ($0.8373$). As a result (see Fig.~\ref{fig:brain}c), it can be sparsified very aggressively without large variations in all measures except for CPL and transitivity. All three mesoscopic group structures: k-means, Louvain and core-periphery have very stable AMI scores. The MASS values also reflect the same trend. 

The coauthor network captures the academic collaborations between institutions revealed through papers authored by Indiana University faculty. It is built from Thomson Reuters’ Web of Science data (Web of Knowledge version 5 \cite{webofSci}) from the 2008 to 2013. Like the airline network, it has a strong core-periphery structure with $0.6247$ weight-degree correlation. However, its peripheral vertices get disconnected at a much slower rate. Again, the mesoscopic group structure remains robust under thresholding, captured by k-means, Louvain, core-periphery, and the MASS curves (see Fig.~\ref{fig:brain}d). In contrast, CPL, DeltaCon, transitivity and global efficiency again demonstrate very different patterns.

We conclude that group structure, including community and core-periphery structures, is a very robust feature that survives even when most edges are removed. The MASS is also quite consistent in capturing group structure across these real world networks. The variation of the measure is rather smooth, and barely affected by the disconnection of small subgraphs, making it a stable and efficient measure for evaluating thresholding effects. \added{We remark that the empirically observed robustness of the group structure does not hold if the weight-degree correlation is destroyed (see Appendix D).}


\added{While we currently lack a full theoretical understanding of the relationship between MASS and community structure, here we provide some mathematical justification.} Matrix perturbation theory studies the change of graph spectrum of a real symmetric matrix $A$ by "perturbing" it. In our context, if we treat the original Laplacian matrix $L$ as $A$ and the difference Laplacian $\varDelta L$ as the perturbation, the classical Weyl’s theorem directly relates to MASS,
\[
  \frac{|\lambda_i - \tilde{\lambda}_i|}{\lambda_N} \le \frac{|\lambda_N^{\varDelta}|}{\lambda_N} = 1-\sigma_{min}, 1\le i\le N\;.
\]

We now consider a generalized version of the Davis-Kahan theorem on subspaces, 
\begin{align*}
\label{eq:davis}
 ||\sin \angle (V_k,\tilde{V}_k)||_F \le \frac{2\sqrt{k}|\lambda_N^{\varDelta}|}{\delta_k}, 1\le k\le N\;,
\end{align*}
where $||A||_F$ represents the Frobenius norm of the matrix $A$, $V_k$, $\tilde{V}_k$ are the $k-dimensional$ subspaces spanned by the eigenbasis $v_1,v_2,...,v_k$ and $\tilde{v}_1, \tilde{v}_2,...,\tilde{v}_k$, respectively, $\delta_k=\lambda_{k+1}-\lambda_k$ and $\sin \angle (V_k,\tilde{V}_k)$ is a diagonal matrix whose entries are the sines of the angles between the corresponding eigenvectors in the two eigenbasis. \added{If $\sigma_{min}$ approaches $1$, or equivalently $\lambda_N^{\varDelta}$ approaches $0$, we then have a tight upper bound on the rotation of the eigenbasis $v_1,v_2,...,v_k$. According to spectral graph theory, these smaller eigenvectors play a fundamental role in defining community structures \cite{sarkar_eigenvector_2015,luxburg_tutorial_2007}. Preservation of MASS therefore guarantees a one sided upper bound on the change to community structure.}

\section{Conclusions}

We have carried out a detailed analysis of weight thresholding on weighted networks.  In general, it appears that group structure is fairly robust under weight thresholding, in contrast to other features. We found that this is due to the peculiar correlation between weight and degree that is commonly observed in real networks, according to which large weights are more likely to be carried by links attached to high degree vertices.

We have also introduced a new measure, the minimum absolute spectral similarity (MASS), to estimate the effect that sparsification procedures have on spectral features of the network. In case studies above we have seen that MASS behaves similarly to traditional group structure measures when there is correlation between weight and degree.

This work deals with weight thresholding, but the analysis can be easily repeated with more sophisticated graph sparsification methods. \added{In the future, we plan to investigate more closely the relationship between MASS and group structure, as well as the role played by the weight-degree correlation.}

\bibliography{references}

\begin{thebibliography}{37}%
\makeatletter
\providecommand \@ifxundefined [1]{%
 \@ifx{#1\undefined}
}%
\providecommand \@ifnum [1]{%
 \ifnum #1\expandafter \@firstoftwo
 \else \expandafter \@secondoftwo
 \fi
}%
\providecommand \@ifx [1]{%
 \ifx #1\expandafter \@firstoftwo
 \else \expandafter \@secondoftwo
 \fi
}%
\providecommand \natexlab [1]{#1}%
\providecommand \enquote  [1]{``#1''}%
\providecommand \bibnamefont  [1]{#1}%
\providecommand \bibfnamefont [1]{#1}%
\providecommand \citenamefont [1]{#1}%
\providecommand \href@noop [0]{\@secondoftwo}%
\providecommand \href [0]{\begingroup \@sanitize@url \@href}%
\providecommand \@href[1]{\@@startlink{#1}\@@href}%
\providecommand \@@href[1]{\endgroup#1\@@endlink}%
\providecommand \@sanitize@url [0]{\catcode `\\12\catcode `\$12\catcode
  `\&12\catcode `\#12\catcode `\^12\catcode `\_12\catcode `\%12\relax}%
\providecommand \@@startlink[1]{}%
\providecommand \@@endlink[0]{}%
\providecommand \url  [0]{\begingroup\@sanitize@url \@url }%
\providecommand \@url [1]{\endgroup\@href {#1}{\urlprefix }}%
\providecommand \urlprefix  [0]{URL }%
\providecommand \Eprint [0]{\href }%
\providecommand \doibase [0]{http://dx.doi.org/}%
\providecommand \selectlanguage [0]{\@gobble}%
\providecommand \bibinfo  [0]{\@secondoftwo}%
\providecommand \bibfield  [0]{\@secondoftwo}%
\providecommand \translation [1]{[#1]}%
\providecommand \BibitemOpen [0]{}%
\providecommand \bibitemStop [0]{}%
\providecommand \bibitemNoStop [0]{.\EOS\space}%
\providecommand \EOS [0]{\spacefactor3000\relax}%
\providecommand \BibitemShut  [1]{\csname bibitem#1\endcsname}%
\let\auto@bib@innerbib\@empty
\bibitem [{\citenamefont {Barrat}\ \emph {et~al.}(2004)\citenamefont {Barrat},
  \citenamefont {Barthelemy}, \citenamefont {Pastor-Satorras},\ and\
  \citenamefont {Vespignani}}]{barrat04}%
  \BibitemOpen
  \bibfield  {author} {\bibinfo {author} {\bibfnamefont {A.}~\bibnamefont
  {Barrat}}, \bibinfo {author} {\bibfnamefont {M.}~\bibnamefont {Barthelemy}},
  \bibinfo {author} {\bibfnamefont {R.}~\bibnamefont {Pastor-Satorras}}, \ and\
  \bibinfo {author} {\bibfnamefont {A.}~\bibnamefont {Vespignani}},\
  }\href@noop {} {\bibfield  {journal} {\bibinfo  {journal} {Proceedings of the
  National Academy of Sciences of the United States of America}\ }\textbf
  {\bibinfo {volume} {101}},\ \bibinfo {pages} {3747} (\bibinfo {year}
  {2004})}\BibitemShut {NoStop}%
\bibitem [{\citenamefont {Namaki}\ \emph {et~al.}(2011)\citenamefont {Namaki},
  \citenamefont {Shirazi}, \citenamefont {Raei},\ and\ \citenamefont
  {Jafari}}]{namaki2011network}%
  \BibitemOpen
  \bibfield  {author} {\bibinfo {author} {\bibfnamefont {A.}~\bibnamefont
  {Namaki}}, \bibinfo {author} {\bibfnamefont {A.}~\bibnamefont {Shirazi}},
  \bibinfo {author} {\bibfnamefont {R.}~\bibnamefont {Raei}}, \ and\ \bibinfo
  {author} {\bibfnamefont {G.}~\bibnamefont {Jafari}},\ }\href@noop {}
  {\bibfield  {journal} {\bibinfo  {journal} {Physica A: Statistical Mechanics
  and its Applications}\ }\textbf {\bibinfo {volume} {390}},\ \bibinfo {pages}
  {3835} (\bibinfo {year} {2011})}\BibitemShut {NoStop}%
\bibitem [{\citenamefont {Bullmore}\ and\ \citenamefont
  {Sporns}(2009)}]{bullmore_complex_2009}%
  \BibitemOpen
  \bibfield  {author} {\bibinfo {author} {\bibfnamefont {E.}~\bibnamefont
  {Bullmore}}\ and\ \bibinfo {author} {\bibfnamefont {O.}~\bibnamefont
  {Sporns}},\ }\href {\doibase 10.1038/nrn2575} {\bibfield  {journal} {\bibinfo
   {journal} {Nat Rev Neurosci}\ }\textbf {\bibinfo {volume} {10}},\ \bibinfo
  {pages} {186} (\bibinfo {year} {2009})}\BibitemShut {NoStop}%
\bibitem [{\citenamefont {Tumminello}\ \emph {et~al.}(2005)\citenamefont
  {Tumminello}, \citenamefont {Aste}, \citenamefont {{Di Matteo}},\ and\
  \citenamefont {Mantegna}}]{tumminello05}%
  \BibitemOpen
  \bibfield  {author} {\bibinfo {author} {\bibfnamefont {M.}~\bibnamefont
  {Tumminello}}, \bibinfo {author} {\bibfnamefont {T.}~\bibnamefont {Aste}},
  \bibinfo {author} {\bibfnamefont {T.}~\bibnamefont {{Di Matteo}}}, \ and\
  \bibinfo {author} {\bibfnamefont {R.~N.}\ \bibnamefont {Mantegna}},\
  }\href@noop {} {\bibfield  {journal} {\bibinfo  {journal} {Proceedings of the
  National Academy of Sciences of the United States of America}\ }\textbf
  {\bibinfo {volume} {102}},\ \bibinfo {pages} {10421} (\bibinfo {year}
  {2005})}\BibitemShut {NoStop}%
\bibitem [{\citenamefont {Serrano}\ \emph {et~al.}(2009)\citenamefont
  {Serrano}, \citenamefont {Bogun{\'a}},\ and\ \citenamefont
  {Vespignani}}]{serrano_extracting_2009}%
  \BibitemOpen
  \bibfield  {author} {\bibinfo {author} {\bibfnamefont {M.~{\'A}.}\
  \bibnamefont {Serrano}}, \bibinfo {author} {\bibfnamefont {M.}~\bibnamefont
  {Bogun{\'a}}}, \ and\ \bibinfo {author} {\bibfnamefont {A.}~\bibnamefont
  {Vespignani}},\ }\href {http://www.pnas.org/content/106/16/6483.short}
  {\bibfield  {journal} {\bibinfo  {journal} {Proceedings of the national
  academy of sciences}\ }\textbf {\bibinfo {volume} {106}},\ \bibinfo {pages}
  {6483} (\bibinfo {year} {2009})}\BibitemShut {NoStop}%
\bibitem [{\citenamefont {Radicchi}\ \emph {et~al.}(2011)\citenamefont
  {Radicchi}, \citenamefont {Ramasco},\ and\ \citenamefont
  {Fortunato}}]{radicchi11}%
  \BibitemOpen
  \bibfield  {author} {\bibinfo {author} {\bibfnamefont {F.}~\bibnamefont
  {Radicchi}}, \bibinfo {author} {\bibfnamefont {J.~J.}\ \bibnamefont
  {Ramasco}}, \ and\ \bibinfo {author} {\bibfnamefont {S.}~\bibnamefont
  {Fortunato}},\ }\href@noop {} {\bibfield  {journal} {\bibinfo  {journal}
  {Physical Review E}\ }\textbf {\bibinfo {volume} {83}},\ \bibinfo {pages}
  {046101} (\bibinfo {year} {2011})}\BibitemShut {NoStop}%
\bibitem [{\citenamefont {Spielman}\ and\ \citenamefont
  {Srivastava}(2011)}]{spielman_resistance_2011}%
  \BibitemOpen
  \bibfield  {author} {\bibinfo {author} {\bibfnamefont {D.~A.}\ \bibnamefont
  {Spielman}}\ and\ \bibinfo {author} {\bibfnamefont {N.}~\bibnamefont
  {Srivastava}},\ }\href {http://epubs.siam.org/doi/abs/10.1137/080734029}
  {\bibfield  {journal} {\bibinfo  {journal} {SIAM Journal on Computing}\
  }\textbf {\bibinfo {volume} {40}},\ \bibinfo {pages} {1913} (\bibinfo {year}
  {2011})}\BibitemShut {NoStop}%
\bibitem [{\citenamefont {Dianati}(2016)}]{Dianati2016}%
  \BibitemOpen
  \bibfield  {author} {\bibinfo {author} {\bibfnamefont {N.}~\bibnamefont
  {Dianati}},\ }\href {\doibase 10.1103/PhysRevE.93.012304} {\bibfield
  {journal} {\bibinfo  {journal} {Phys. Rev. E}\ }\textbf {\bibinfo {volume}
  {93}},\ \bibinfo {pages} {012304} (\bibinfo {year} {2016})}\BibitemShut
  {NoStop}%
\bibitem [{\citenamefont {Coscia}\ and\ \citenamefont
  {Neffke}(2017)}]{coscia2017backboning}%
  \BibitemOpen
  \bibfield  {author} {\bibinfo {author} {\bibfnamefont {M.}~\bibnamefont
  {Coscia}}\ and\ \bibinfo {author} {\bibfnamefont {F.~M.~H.}\ \bibnamefont
  {Neffke}},\ }in\ \href {\doibase 10.1109/ICDE.2017.100} {\emph {\bibinfo
  {booktitle} {2017 IEEE 33rd International Conference on Data Engineering
  (ICDE)}}}\ (\bibinfo {year} {2017})\ pp.\ \bibinfo {pages}
  {425--436}\BibitemShut {NoStop}%
\bibitem [{\citenamefont {{Kobayashi}}\ \emph {et~al.}(2018)\citenamefont
  {{Kobayashi}}, \citenamefont {{Takaguchi}},\ and\ \citenamefont
  {{Barrat}}}]{2018arXivKobayashi}%
  \BibitemOpen
  \bibfield  {author} {\bibinfo {author} {\bibfnamefont {T.}~\bibnamefont
  {{Kobayashi}}}, \bibinfo {author} {\bibfnamefont {T.}~\bibnamefont
  {{Takaguchi}}}, \ and\ \bibinfo {author} {\bibfnamefont {A.}~\bibnamefont
  {{Barrat}}},\ }\href@noop {} {\bibfield  {journal} {\bibinfo  {journal}
  {ArXiv e-prints}\ } (\bibinfo {year} {2018})},\ \Eprint
  {http://arxiv.org/abs/1804.08828} {arXiv:1804.08828 [physics.soc-ph]}
  \BibitemShut {NoStop}%
\bibitem [{\citenamefont {Lynall}\ \emph {et~al.}(2010)\citenamefont {Lynall},
  \citenamefont {Bassett}, \citenamefont {Kerwin}, \citenamefont {McKenna},
  \citenamefont {Kitzbichler}, \citenamefont {M{\"u}ller},\ and\ \citenamefont
  {Bullmore}}]{lynall_functional_2010}%
  \BibitemOpen
  \bibfield  {author} {\bibinfo {author} {\bibfnamefont {M.-E.}\ \bibnamefont
  {Lynall}}, \bibinfo {author} {\bibfnamefont {D.~S.}\ \bibnamefont {Bassett}},
  \bibinfo {author} {\bibfnamefont {R.}~\bibnamefont {Kerwin}}, \bibinfo
  {author} {\bibfnamefont {P.~J.}\ \bibnamefont {McKenna}}, \bibinfo {author}
  {\bibfnamefont {M.}~\bibnamefont {Kitzbichler}}, \bibinfo {author}
  {\bibfnamefont {U.}~\bibnamefont {M{\"u}ller}}, \ and\ \bibinfo {author}
  {\bibfnamefont {E.}~\bibnamefont {Bullmore}},\ }\href {\doibase
  10.1523/JNEUROSCI.0333-10.2010} {\bibfield  {journal} {\bibinfo  {journal}
  {The Journal of neuroscience : the official journal of the Society for
  Neuroscience}\ }\textbf {\bibinfo {volume} {30}},\ \bibinfo {pages} {9477}
  (\bibinfo {year} {2010})}\BibitemShut {NoStop}%
\bibitem [{\citenamefont {Allesina}\ \emph {et~al.}(2006)\citenamefont
  {Allesina}, \citenamefont {Bodini},\ and\ \citenamefont
  {Bondavalli}}]{allesina_secondary_2006}%
  \BibitemOpen
  \bibfield  {author} {\bibinfo {author} {\bibfnamefont {S.}~\bibnamefont
  {Allesina}}, \bibinfo {author} {\bibfnamefont {A.}~\bibnamefont {Bodini}}, \
  and\ \bibinfo {author} {\bibfnamefont {C.}~\bibnamefont {Bondavalli}},\
  }\href {\doibase 10.1016/j.ecolmodel.2005.10.016} {\bibfield  {journal}
  {\bibinfo  {journal} {Ecological Modelling}\ }\textbf {\bibinfo {volume}
  {194}},\ \bibinfo {pages} {150} (\bibinfo {year} {2006})}\BibitemShut
  {NoStop}%
\bibitem [{\citenamefont {Garrison}\ \emph {et~al.}(2015)\citenamefont
  {Garrison}, \citenamefont {Scheinost}, \citenamefont {Finn}, \citenamefont
  {Shen},\ and\ \citenamefont {Constable}}]{garrison_stability_2015}%
  \BibitemOpen
  \bibfield  {author} {\bibinfo {author} {\bibfnamefont {K.~A.}\ \bibnamefont
  {Garrison}}, \bibinfo {author} {\bibfnamefont {D.}~\bibnamefont {Scheinost}},
  \bibinfo {author} {\bibfnamefont {E.~S.}\ \bibnamefont {Finn}}, \bibinfo
  {author} {\bibfnamefont {X.}~\bibnamefont {Shen}}, \ and\ \bibinfo {author}
  {\bibfnamefont {R.~T.}\ \bibnamefont {Constable}},\ }\href {\doibase
  10.1016/j.neuroimage.2015.05.046} {\bibfield  {journal} {\bibinfo  {journal}
  {NeuroImage}\ }\textbf {\bibinfo {volume} {118}},\ \bibinfo {pages} {651}
  (\bibinfo {year} {2015})}\BibitemShut {NoStop}%
\bibitem [{\citenamefont {Fortunato}(2010)}]{fortunato2010community}%
  \BibitemOpen
  \bibfield  {author} {\bibinfo {author} {\bibfnamefont {S.}~\bibnamefont
  {Fortunato}},\ }\href@noop {} {\bibfield  {journal} {\bibinfo  {journal}
  {Physics reports}\ }\textbf {\bibinfo {volume} {486}},\ \bibinfo {pages} {75}
  (\bibinfo {year} {2010})}\BibitemShut {NoStop}%
\bibitem [{\citenamefont {Luxburg}(2007)}]{luxburg_tutorial_2007}%
  \BibitemOpen
  \bibfield  {author} {\bibinfo {author} {\bibfnamefont {U.}~\bibnamefont
  {Luxburg}},\ }\href {\doibase 10.1007/s11222-007-9033-z} {\bibfield
  {journal} {\bibinfo  {journal} {Statistics and Computing}\ }\textbf {\bibinfo
  {volume} {17}},\ \bibinfo {pages} {395} (\bibinfo {year} {2007})}\BibitemShut
  {NoStop}%
\bibitem [{\citenamefont {Sarkar}\ \emph {et~al.}(2015)\citenamefont {Sarkar},
  \citenamefont {Chawla}, \citenamefont {Robinson},\ and\ \citenamefont
  {Fortunato}}]{sarkar_eigenvector_2015}%
  \BibitemOpen
  \bibfield  {author} {\bibinfo {author} {\bibfnamefont {S.}~\bibnamefont
  {Sarkar}}, \bibinfo {author} {\bibfnamefont {S.}~\bibnamefont {Chawla}},
  \bibinfo {author} {\bibfnamefont {P.~A.}\ \bibnamefont {Robinson}}, \ and\
  \bibinfo {author} {\bibfnamefont {S.}~\bibnamefont {Fortunato}},\ }\href
  {http://arxiv.org/abs/1510.07064} {\bibfield  {journal} {\bibinfo  {journal}
  {arXiv preprint arXiv:1510.07064}\ } (\bibinfo {year} {2015})}\BibitemShut
  {NoStop}%
\bibitem [{\citenamefont {Iriarte}(2016)}]{Iriarte2016}%
  \BibitemOpen
  \bibfield  {author} {\bibinfo {author} {\bibfnamefont {B.}~\bibnamefont
  {Iriarte}},\ }\href {\doibase 10.1137/15M1008737} {\bibfield  {journal}
  {\bibinfo  {journal} {SIAM Journal on Discrete Mathematics}\ }\textbf
  {\bibinfo {volume} {30}},\ \bibinfo {pages} {2146} (\bibinfo {year}
  {2016})},\ \Eprint {http://arxiv.org/abs/https://doi.org/10.1137/15M1008737}
  {https://doi.org/10.1137/15M1008737} \BibitemShut {NoStop}%
\bibitem [{\citenamefont {{Lancichinetti}}\ \emph {et~al.}(2008)\citenamefont
  {{Lancichinetti}}, \citenamefont {{Fortunato}},\ and\ \citenamefont
  {{Radicchi}}}]{2008LFR}%
  \BibitemOpen
  \bibfield  {author} {\bibinfo {author} {\bibfnamefont {A.}~\bibnamefont
  {{Lancichinetti}}}, \bibinfo {author} {\bibfnamefont {S.}~\bibnamefont
  {{Fortunato}}}, \ and\ \bibinfo {author} {\bibfnamefont {F.}~\bibnamefont
  {{Radicchi}}},\ }\href {\doibase 10.1103/PhysRevE.78.046110} {\bibfield
  {journal} {\bibinfo  {journal} {\pre}\ }\textbf {\bibinfo {volume} {78}},\
  \bibinfo {eid} {046110} (\bibinfo {year} {2008})},\ \Eprint
  {http://arxiv.org/abs/0805.4770} {arXiv:0805.4770 [physics.soc-ph]}
  \BibitemShut {NoStop}%
\bibitem [{\citenamefont {Betzel}\ \emph {et~al.}(2013)\citenamefont {Betzel},
  \citenamefont {Griffa}, \citenamefont {Avena-Koenigsberger}, \citenamefont
  {Go{\~n}i}, \citenamefont {Thiran}, \citenamefont {Hagmann},\ and\
  \citenamefont {Sporns}}]{betzel2013multi}%
  \BibitemOpen
  \bibfield  {author} {\bibinfo {author} {\bibfnamefont {R.~F.}\ \bibnamefont
  {Betzel}}, \bibinfo {author} {\bibfnamefont {A.}~\bibnamefont {Griffa}},
  \bibinfo {author} {\bibfnamefont {A.}~\bibnamefont {Avena-Koenigsberger}},
  \bibinfo {author} {\bibfnamefont {J.}~\bibnamefont {Go{\~n}i}}, \bibinfo
  {author} {\bibfnamefont {J.-P.}\ \bibnamefont {Thiran}}, \bibinfo {author}
  {\bibfnamefont {P.}~\bibnamefont {Hagmann}}, \ and\ \bibinfo {author}
  {\bibfnamefont {O.}~\bibnamefont {Sporns}},\ }\href@noop {} {\bibfield
  {journal} {\bibinfo  {journal} {Network Science}\ }\textbf {\bibinfo {volume}
  {1}},\ \bibinfo {pages} {353} (\bibinfo {year} {2013})}\BibitemShut {NoStop}%
\bibitem [{\citenamefont {{De Benedictis}}\ and\ \citenamefont
  {Tajoli}(2011)}]{de2011trade}%
  \BibitemOpen
  \bibfield  {author} {\bibinfo {author} {\bibfnamefont {L.}~\bibnamefont {{De
  Benedictis}}}\ and\ \bibinfo {author} {\bibfnamefont {L.}~\bibnamefont
  {Tajoli}},\ }\href@noop {} {\bibfield  {journal} {\bibinfo  {journal} {The
  World Economy}\ }\textbf {\bibinfo {volume} {34}},\ \bibinfo {pages} {1417}
  (\bibinfo {year} {2011})}\BibitemShut {NoStop}%
\bibitem [{\citenamefont {Patokallio}(2009)}]{openFlights}%
  \BibitemOpen
  \bibfield  {author} {\bibinfo {author} {\bibfnamefont {J.}~\bibnamefont
  {Patokallio}},\ }\href@noop {} {\bibfield  {journal} {\bibinfo  {journal}
  {{http://openflights.org/}}\ } (\bibinfo {year} {2009})}\BibitemShut
  {NoStop}%
\bibitem [{\citenamefont {Latora}\ and\ \citenamefont
  {Marchiori}(2001)}]{Latora2001Efficient}%
  \BibitemOpen
  \bibfield  {author} {\bibinfo {author} {\bibfnamefont {V.}~\bibnamefont
  {Latora}}\ and\ \bibinfo {author} {\bibfnamefont {M.}~\bibnamefont
  {Marchiori}},\ }\href {\doibase 10.1103/PhysRevLett.87.198701} {\bibfield
  {journal} {\bibinfo  {journal} {Phys. Rev. Lett.}\ }\textbf {\bibinfo
  {volume} {87}},\ \bibinfo {pages} {198701} (\bibinfo {year}
  {2001})}\BibitemShut {NoStop}%
\bibitem [{\citenamefont {Newman}(2006)}]{newman_modularity_2006}%
  \BibitemOpen
  \bibfield  {author} {\bibinfo {author} {\bibfnamefont {M.~E.~J.}\
  \bibnamefont {Newman}},\ }\href@noop {} {\bibfield  {journal} {\bibinfo
  {journal} {Proceedings of the National Academy of Sciences}\ }\textbf
  {\bibinfo {volume} {103}},\ \bibinfo {pages} {8577} (\bibinfo {year}
  {2006})}\BibitemShut {NoStop}%
\bibitem [{\citenamefont {{Blondel}}\ \emph {et~al.}(2008)\citenamefont
  {{Blondel}}, \citenamefont {{Guillaume}}, \citenamefont {{Lambiotte}},\ and\
  \citenamefont {{Lefebvre}}}]{Blondel2008Louvain}%
  \BibitemOpen
  \bibfield  {author} {\bibinfo {author} {\bibfnamefont {V.~D.}\ \bibnamefont
  {{Blondel}}}, \bibinfo {author} {\bibfnamefont {J.-L.}\ \bibnamefont
  {{Guillaume}}}, \bibinfo {author} {\bibfnamefont {R.}~\bibnamefont
  {{Lambiotte}}}, \ and\ \bibinfo {author} {\bibfnamefont {E.}~\bibnamefont
  {{Lefebvre}}},\ }\href {\doibase 10.1088/1742-5468/2008/10/P10008} {\bibfield
   {journal} {\bibinfo  {journal} {Journal of Statistical Mechanics: Theory and
  Experiment}\ }\textbf {\bibinfo {volume} {10}},\ \bibinfo {pages} {10008}
  (\bibinfo {year} {2008})},\ \Eprint {http://arxiv.org/abs/0803.0476}
  {arXiv:0803.0476 [physics.soc-ph]} \BibitemShut {NoStop}%
\bibitem [{\citenamefont {Mucha}\ \emph {et~al.}(2010)\citenamefont {Mucha},
  \citenamefont {Richardson}, \citenamefont {Macon}, \citenamefont {Porter},\
  and\ \citenamefont {Onnela}}]{mucha_community_2010}%
  \BibitemOpen
  \bibfield  {author} {\bibinfo {author} {\bibfnamefont {P.~J.}\ \bibnamefont
  {Mucha}}, \bibinfo {author} {\bibfnamefont {T.}~\bibnamefont {Richardson}},
  \bibinfo {author} {\bibfnamefont {K.}~\bibnamefont {Macon}}, \bibinfo
  {author} {\bibfnamefont {M.~A.}\ \bibnamefont {Porter}}, \ and\ \bibinfo
  {author} {\bibfnamefont {J.-P.}\ \bibnamefont {Onnela}},\ }\href {\doibase
  10.1126/science.1184819} {\bibfield  {journal} {\bibinfo  {journal}
  {Science}\ }\textbf {\bibinfo {volume} {328}},\ \bibinfo {pages} {876}
  (\bibinfo {year} {2010})}\BibitemShut {NoStop}%
\bibitem [{\citenamefont {Rubinov}\ \emph {et~al.}(2015)\citenamefont
  {Rubinov}, \citenamefont {Ypma}, \citenamefont {Watson},\ and\ \citenamefont
  {Bullmore}}]{Rubinov2015}%
  \BibitemOpen
  \bibfield  {author} {\bibinfo {author} {\bibfnamefont {M.}~\bibnamefont
  {Rubinov}}, \bibinfo {author} {\bibfnamefont {R.~J.~F.}\ \bibnamefont
  {Ypma}}, \bibinfo {author} {\bibfnamefont {C.}~\bibnamefont {Watson}}, \ and\
  \bibinfo {author} {\bibfnamefont {E.~T.}\ \bibnamefont {Bullmore}},\ }\href
  {\doibase 10.1073/pnas.1420315112} {\bibfield  {journal} {\bibinfo  {journal}
  {Proceedings of the National Academy of Sciences}\ }\textbf {\bibinfo
  {volume} {112}},\ \bibinfo {pages} {10032} (\bibinfo {year} {2015})},\
  \Eprint
  {http://arxiv.org/abs/http://www.pnas.org/content/112/32/10032.full.pdf}
  {http://www.pnas.org/content/112/32/10032.full.pdf} \BibitemShut {NoStop}%
\bibitem [{\citenamefont {Koutra}\ \emph {et~al.}(2016)\citenamefont {Koutra},
  \citenamefont {Shah}, \citenamefont {Vogelstein}, \citenamefont {Gallagher},\
  and\ \citenamefont {Faloutsos}}]{Koutra_DeltaCon_2016}%
  \BibitemOpen
  \bibfield  {author} {\bibinfo {author} {\bibfnamefont {D.}~\bibnamefont
  {Koutra}}, \bibinfo {author} {\bibfnamefont {N.}~\bibnamefont {Shah}},
  \bibinfo {author} {\bibfnamefont {J.~T.}\ \bibnamefont {Vogelstein}},
  \bibinfo {author} {\bibfnamefont {B.}~\bibnamefont {Gallagher}}, \ and\
  \bibinfo {author} {\bibfnamefont {C.}~\bibnamefont {Faloutsos}},\ }\href
  {\doibase 10.1145/2824443} {\bibfield  {journal} {\bibinfo  {journal} {ACM
  Trans. Knowl. Discov. Data}\ }\textbf {\bibinfo {volume} {10}},\ \bibinfo
  {pages} {28:1} (\bibinfo {year} {2016})}\BibitemShut {NoStop}%
\bibitem [{\citenamefont {Vinh}\ \emph {et~al.}(2010)\citenamefont {Vinh},
  \citenamefont {Epps},\ and\ \citenamefont {Bailey}}]{vinh2010information}%
  \BibitemOpen
  \bibfield  {author} {\bibinfo {author} {\bibfnamefont {N.~X.}\ \bibnamefont
  {Vinh}}, \bibinfo {author} {\bibfnamefont {J.}~\bibnamefont {Epps}}, \ and\
  \bibinfo {author} {\bibfnamefont {J.}~\bibnamefont {Bailey}},\ }\href@noop {}
  {\bibfield  {journal} {\bibinfo  {journal} {Journal of Machine Learning
  Research}\ }\textbf {\bibinfo {volume} {11}},\ \bibinfo {pages} {2837}
  (\bibinfo {year} {2010})}\BibitemShut {NoStop}%
\bibitem [{\citenamefont {Spielman}\ and\ \citenamefont
  {Teng}(2011)}]{spielman_spectral_2008}%
  \BibitemOpen
  \bibfield  {author} {\bibinfo {author} {\bibfnamefont {D.~A.}\ \bibnamefont
  {Spielman}}\ and\ \bibinfo {author} {\bibfnamefont {S.-H.}\ \bibnamefont
  {Teng}},\ }\href@noop {} {\bibfield  {journal} {\bibinfo  {journal} {SIAM
  Journal on Computing}\ }\textbf {\bibinfo {volume} {40}},\ \bibinfo {pages}
  {981} (\bibinfo {year} {2011})}\BibitemShut {NoStop}%
\bibitem [{\citenamefont {Sorensen}(1992)}]{sorensen1992implicit}%
  \BibitemOpen
  \bibfield  {author} {\bibinfo {author} {\bibfnamefont {D.~C.}\ \bibnamefont
  {Sorensen}},\ }\href@noop {} {\bibfield  {journal} {\bibinfo  {journal} {Siam
  journal on matrix analysis and applications}\ }\textbf {\bibinfo {volume}
  {13}},\ \bibinfo {pages} {357} (\bibinfo {year} {1992})}\BibitemShut
  {NoStop}%
\bibitem [{\citenamefont {Lehoucq}\ \emph {et~al.}(1998)\citenamefont
  {Lehoucq}, \citenamefont {Sorensen},\ and\ \citenamefont
  {Yang}}]{1998_arpack}%
  \BibitemOpen
  \bibfield  {author} {\bibinfo {author} {\bibfnamefont {R.~B.}\ \bibnamefont
  {Lehoucq}}, \bibinfo {author} {\bibfnamefont {D.~C.}\ \bibnamefont
  {Sorensen}}, \ and\ \bibinfo {author} {\bibfnamefont {C.}~\bibnamefont
  {Yang}},\ }\href@noop {} {\emph {\bibinfo {title} {{ARPACK users' guide:
  solution of large-scale eigenvalue problems with implicitly restarted Arnoldi
  methods}}}}\ (\bibinfo  {publisher} {SIAM},\ \bibinfo {year}
  {1998})\BibitemShut {NoStop}%
\bibitem [{\citenamefont {Stewart}(2001)}]{Stewart_2001}%
  \BibitemOpen
  \bibfield  {author} {\bibinfo {author} {\bibfnamefont {G.~W.}\ \bibnamefont
  {Stewart}},\ }\href {\doibase 10.1137/S0895479800371529} {\bibfield
  {journal} {\bibinfo  {journal} {SIAM J. Matrix Anal. Appl.}\ }\textbf
  {\bibinfo {volume} {23}},\ \bibinfo {pages} {601} (\bibinfo {year}
  {2001})}\BibitemShut {NoStop}%
\bibitem [{\citenamefont {{Lancichinetti}}\ and\ \citenamefont
  {{Fortunato}}(2009)}]{2009LFweighted}%
  \BibitemOpen
  \bibfield  {author} {\bibinfo {author} {\bibfnamefont {A.}~\bibnamefont
  {{Lancichinetti}}}\ and\ \bibinfo {author} {\bibfnamefont {S.}~\bibnamefont
  {{Fortunato}}},\ }\href@noop {} {\bibfield  {journal} {\bibinfo  {journal}
  {ArXiv e-prints}\ } (\bibinfo {year} {2009})},\ \Eprint
  {http://arxiv.org/abs/0904.3940} {arXiv:0904.3940 [physics.soc-ph]}
  \BibitemShut {NoStop}%
\bibitem [{\citenamefont {{Clarivate Analytics}}(2015)}]{webofSci}%
  \BibitemOpen
  \bibfield  {author} {\bibinfo {author} {\bibfnamefont {T.~R.}\ \bibnamefont
  {{Clarivate Analytics}}},\ }\href@noop {} {\bibfield  {journal} {\bibinfo
  {journal} {{http://iuni.iu.edu/resources/web-of-science}}\ } (\bibinfo {year}
  {2015})}\BibitemShut {NoStop}%
\bibitem [{\citenamefont {Rubinov}\ and\ \citenamefont
  {Sporns}(2010)}]{Rubinov2010Sporns}%
  \BibitemOpen
  \bibfield  {author} {\bibinfo {author} {\bibfnamefont {M.}~\bibnamefont
  {Rubinov}}\ and\ \bibinfo {author} {\bibfnamefont {O.}~\bibnamefont
  {Sporns}},\ }\href {\doibase 10.1016/j.neuroimage.2009.10.003} {\bibfield
  {journal} {\bibinfo  {journal} {NeuroImage}\ }\textbf {\bibinfo {volume}
  {52}},\ \bibinfo {pages} {1059} (\bibinfo {year} {2010})},\ \bibinfo {note}
  {computational Models of the Brain}\BibitemShut {NoStop}%
\bibitem [{\citenamefont {Jeub}\ \emph {et~al.}(2017)\citenamefont {Jeub},
  \citenamefont {Bazzi}, \citenamefont {Jutla},\ and\ \citenamefont
  {Mucha}}]{genLouvain}%
  \BibitemOpen
  \bibfield  {author} {\bibinfo {author} {\bibfnamefont {L.~G.~S.}\
  \bibnamefont {Jeub}}, \bibinfo {author} {\bibfnamefont {M.}~\bibnamefont
  {Bazzi}}, \bibinfo {author} {\bibfnamefont {I.~S.}\ \bibnamefont {Jutla}}, \
  and\ \bibinfo {author} {\bibfnamefont {P.~J.}\ \bibnamefont {Mucha}},\
  }\href@noop {} {\enquote {\bibinfo {title} {{A generalized Louvain method for
  community detection implemented in MATLAB.
  http://netwiki.amath.unc.edu/GenLouvain}},}\ } (\bibinfo {year}
  {2011-2017})\BibitemShut {NoStop}%
\bibitem [{\citenamefont {Haveliwala}(2003)}]{haveliwala2003topic}%
  \BibitemOpen
  \bibfield  {author} {\bibinfo {author} {\bibfnamefont {T.~H.}\ \bibnamefont
  {Haveliwala}},\ }\href@noop {} {\bibfield  {journal} {\bibinfo  {journal}
  {IEEE transactions on knowledge and data engineering}\ }\textbf {\bibinfo
  {volume} {15}},\ \bibinfo {pages} {784} (\bibinfo {year} {2003})}\BibitemShut
  {NoStop}%
\end{thebibliography}%

\appendix
\counterwithin{figure}{section}
\counterwithin{table}{section}

\section{Algorithmic details of graph properties}
Here we provide more details of how the graph properties are defined and calculated. All experiments are conducted under Matlab version R2017a.

{\it Characteristic path length (CPL)} is the average length of all shortest paths connecting pairs of vertices of the network. Once the network becomes disconnected, CPL is not defined, and we set its value to $0$.
 
{\it Transitivity}, or global clustering coefficient, is the ratio of triangles to triplets in the network, where a triplet is a motif consisting of one vertex and two links incident to the vertex.

{\it Global efficiency} is defined as $\frac{2}{N(N-1)}\sum_{u,v}\frac{1}{d_{uv}}$, where $d_{uv}$ represents the shortest path between the vertices $u$ and $v$ \cite{Latora2001Efficient}. Notice here if we define $d_{uv}=\infty$ for unreachable vertex pairs, global efficiency remains robust under network disconnections. The Matlab code for CPL, transitivity and global efficiency are provided by the Brain Connectivity Toolbox \cite{Rubinov2010Sporns}.

{\it Community structure}: The specific Louvain implementation we used is \cite{genLouvain} with all default parameter settings, whereas the $k$-means spectral clustering algorithm follows the pseudo-code in \cite{luxburg_tutorial_2007} with a normalized graph Laplacian. The constant $k$ that is set to be the planted ground truth in synthetic experiments. For real world networks, we set it to be the same stable $k$ value found by the Louvain algorithm (For all 4 networks Louvain was able to find stable $k$s). 

{\it Core-periphery structure}: The partition of core and periphery vertices uses the Matlab code provide by the authors of \cite{Rubinov2015}. The algorithm is based on the following definition of the weighted coreness measure,
\begin{equation}
\label{eq:coreness}
 Q_C = \frac{1}{Z} \left(\sum_{u,v\in C_c}(W_{uv}-\bar{W}) - \sum_{u,v\in C_p}(W_{uv}-\bar{W}) \right)\;.
\end{equation}

In Eq. \ref{eq:coreness}, $C_c$ and $C_p$ represents the bisection of the network into the core and periphery subsets, $\bar{W}$ is the average edge weight and $Z=\sum_{u,v} |W_{uv}-\bar{W}|$ is the normalizing constant, so that $0\le Q_C\le 1$. The idea is that if there is a core-periphery structure, there are many heavy edges joining pairs of vertices of the core $C_c$ and many light edges joining pairs of vertices of the periphery, yielding a value of $Q_C$ appreciably larger than $0$. Therefore, by maximizing $Q^{max}_C$ over all possible bisections of the network, we also get the $C_c$ and $C_p$ for our experiments.

We measure the similarity between the partitions of the sparsified graph $\tilde{G}$ and those of $G$ using the {\it adjusted mutual information} (AMI)~\cite{vinh2010information}, for all three partitioning algorithms, : Louvain, $k$-means and coreness. Notice here that the AMI is taken after removal of single isolated vertices, because they do not constitute meaningful communities. To overcome the randomness of the partitioning algorithms, we sample 100 different partition outcomes from the original graph, 10 from the sparsified graph, and use the maximum AMI between any pair. These numbers are picked so that the same algorithm returns consistent results (AMI very close to 1) on independent runs on the original graph. The resulting curve is thus an upper bound of any individual pairs.
 
{\it DeltaCon}~\cite{Koutra_DeltaCon_2016} is a general graph similarity metric based on diffusion results on graphs. It aggregates the affinities between all pairs of vertices using a rooted Euclidean distance (RootED), 
\[RootED(S_{G}, S_{\tilde{G}}) = \sqrt{\sum_{u,v}(\sqrt{S_{G}^{uv}}-\sqrt{S_{\tilde{G}}^{uv}})},
\]
where the vertex affinity matrices $S_{G}$ and $S_{\tilde{G}}$ are calculated by distributed diffusion processes around each vertex. The affinity matrix involves the calculation of personalized PageRank \cite{haveliwala2003topic}, which is intuitively captured by the following matrix power series,
\[S_{G} \approx [I-\epsilon W]^{-1} = I + \epsilon W + \epsilon^2 W^2 + \epsilon^3 W^3 + ...
\]
Here, $0\le \epsilon\le 1$ represents the decay factor of diffusion over longer distances. Under the default settings, with $\epsilon = \frac{1}{\max_u D_{u}}$, longer range diffusion decays quickly, and DeltaCon thus puts a stronger emphasis on local structure. We use the Matlab code provided by the authors of \cite{Koutra_DeltaCon_2016}, which uses a fast Belief Propagation approximation of personalized PageRank. Since DeltaCon scales directly with weights, we normalize the matrix by multiplying a constant so that DeltaCon reaches $0$ on empty graphs. 

For the LFR benchmark \cite{2008LFR}, we used the binary version downloaded from "https://sites.google.com /site/andrealancichinetti/files". We generated multiple synthetic networks until we get one with $8$ planted communities. The parameters of the LFR benchmark are listed in Table \ref{tab:LFR}.

\begin{table*}
\caption{Parameters of the LFR benchmark}
\begin{minipage}{\textwidth}\begin{center}
\setlength{\tabcolsep}{.3em}
\bgroup\def\arraystretch{1.5}
\begin{tabular}{lc}
	\hline\hline
	Number of vertices  	   &$1000$ \\
	Average degree  	   &$20$     \\ 
	Maximum degree         &$100$     \\
	Minus exponent for the degree sequence       &$2$     \\
	Minus exponent for the community size distribution         &$1$      \\
	Minimum community size         &$10$     \\
	Maximum community size         &$800$     \\
	\hline\hline
	\end{tabular}
	\egroup
\end{center}\end{minipage}
\label{tab:LFR}
\end{table*}

\section{Optimality of linear threshold under expected spectral similarity}
In \eqref{eq:ass}, we defined the \emph{absolute spectral similarity} as a function of the input vector $x$. Besides the worst case $\sigma_{min}$, we can also define an average case similarity measure,
\begin{align*}
 \sigma_{exp} =& E_{x\sim P(x)}\left[1 - \dfrac{x^T\varDelta Lx}{\lambda_N}\right] =  1 -  \dfrac{E_{x\sim P(x)}[x^T\varDelta Lx]}{\lambda_N}\;,
\end{align*}
where the input vectors are drawn from a distribution $P(x)$. If we assume the entries of the vector $x$ are independent identically distributed random variables, we have
\begin{align}
\label{eq:expectedSS}
 \sigma_{exp} =& 1 - \dfrac{\sum_{u\in V, v\in V} \varDelta W_{uv}E_{x_u,x_v\sim P(x)}\left[(x_u-x_v)^2\right]}{\lambda_N}\nonumber\\
             \propto&  1- \dfrac{C}{\lambda_N}\sum_{u\in V, v\in V} (W_{uv}-\tilde{W}_{uv})\;,
\end{align}
where we have used linearity of expectation and the fact that $C=E_{x_u,x_v\sim P(x)}\left[(x_u-x_v)^2\right]$ is independent of $u$ and $v$ because the entries of $x$ are assumed to be independent and identically distributed. Hence, the \emph{expected spectral similarity} $\sigma_{exp}$ simply becomes the maximum when total edge weights are kept as much as possible. This completes the proof that weight thresholding optimizes the expected value of $\sigma(x)$ with independent identically distributed $x$ entries.

\section{Computational efficiency of MASS}
Recall that the MASS measure is defined as
\begin{eqnarray}
\label{eq:minSS2}
 \sigma_{min} &=& \min_{\forall x, |x| = 1}(1 - \dfrac{x^T\varDelta Lx}{\lambda_N}) = 1- \dfrac{\lambda_N^{\varDelta}}{\lambda_N} \;,
\end{eqnarray}
where $\lambda_N^{\varDelta}$ is the largest eigenvalue of the difference Laplacian $\varDelta L$. 

As a general spectral measure, MASS automatically captures important mesoscopic structures in the data. We suggest users of MASS take all types of group structure in to account. However, if the application really concerns the community structures, additional validation can be done relatively easily. The Davis-Kahan theorem (see the main text) provides a theoretical connection between the largest eigenvalue of the difference Laplacian and the smallest eigenvectors associated with community structures.  Empirically, we can also consider the \emph{average rotational angle} (ARA) of the corresponding eigenvectors,
\begin{align}
\label{eq:control}
 \eta = E_{2\le i\le k} [\cos \angle (v_i,\tilde{v_i})]\;,
\end{align}
where $\cos \angle (v_i,\tilde{v_i})$ represents the cosine of the angle between the respective eigenvectors of $G$ and $\tilde{G}$ corresponding to their $i$-th smallest eigenvalue. The integer $k$ is the number of relevant communities and can be selected based on spectral graph theory, the Louvain method or domain knowledge if it is available. 

A major advantage of the formulation in Eq. \eqref{eq:minSS2} is its numerical stability and computational efficiency. Designing efficient and stable algorithms for finding the eigenvalues of a matrix is one of the most important problems in numerical analysis. The state-of-the-art iterative solvers in popular numerical packages today are inherently more stable for larger eigenvalues, including the current implementation of Matlab which we use for this work \cite{sorensen1992implicit,1998_arpack,Stewart_2001}. \added{Readers can find our Matlab implementation at https://github.com/IU-AMBITION/MASS.}

With only computing the largest eigenvalues of two graph Laplacians, the MASS measure is therefore among the most efficient and stable spectral properties. To demonstrate its computational efficiency, we compare the running time of computing MASS [Eqs. \eqref{eq:minSS}] with those of traditional community detection algorithms, as well as DeltaCon and efficiency measures in the Table~\ref{tab:time}. 

\begin{table*}
\caption{Computational time (in seconds) of graph properties}
\begin{minipage}{\textwidth}\begin{center}
\setlength{\tabcolsep}{.3em}
\bgroup\def\arraystretch{1.5}
	\begin{tabular}{lcccccc}
	\hline\hline
	\emph{Measures} 	& \emph{MASS}  &\emph{DeltaCon}  &\emph{Global efficiency} & \emph{K-means} & \emph{Louvain}\\
	\hline
	World trade network  	   &0.181   	&0.457		&154.5   &1.619 &1.355\\
	Airline network   	   &13.60   	&767.2		&157.0   &37.63 &112.9\\ 
	\hline\hline
	\end{tabular}\\
\egroup
\begin{footnotesize}All measures are taken $21$ times, in correspondence to the thresholds $[0,0.05,0.1,0.15,...,0.95,1]$.\\ The experiment is conducted under Matlab version R2017a. Code packages are provided by the authors of \cite{genLouvain,Rubinov2010Sporns,Koutra_DeltaCon_2016}.\end{footnotesize}
\end{center}\end{minipage}
\label{tab:time}
\end{table*}

\section{Results on networks with randomized edge weight}
\begin{figure*}
\centering
 \includegraphics[width=0.95\linewidth]{legend2-eps-converted-to}\\
{\includegraphics[width=0.48\linewidth]{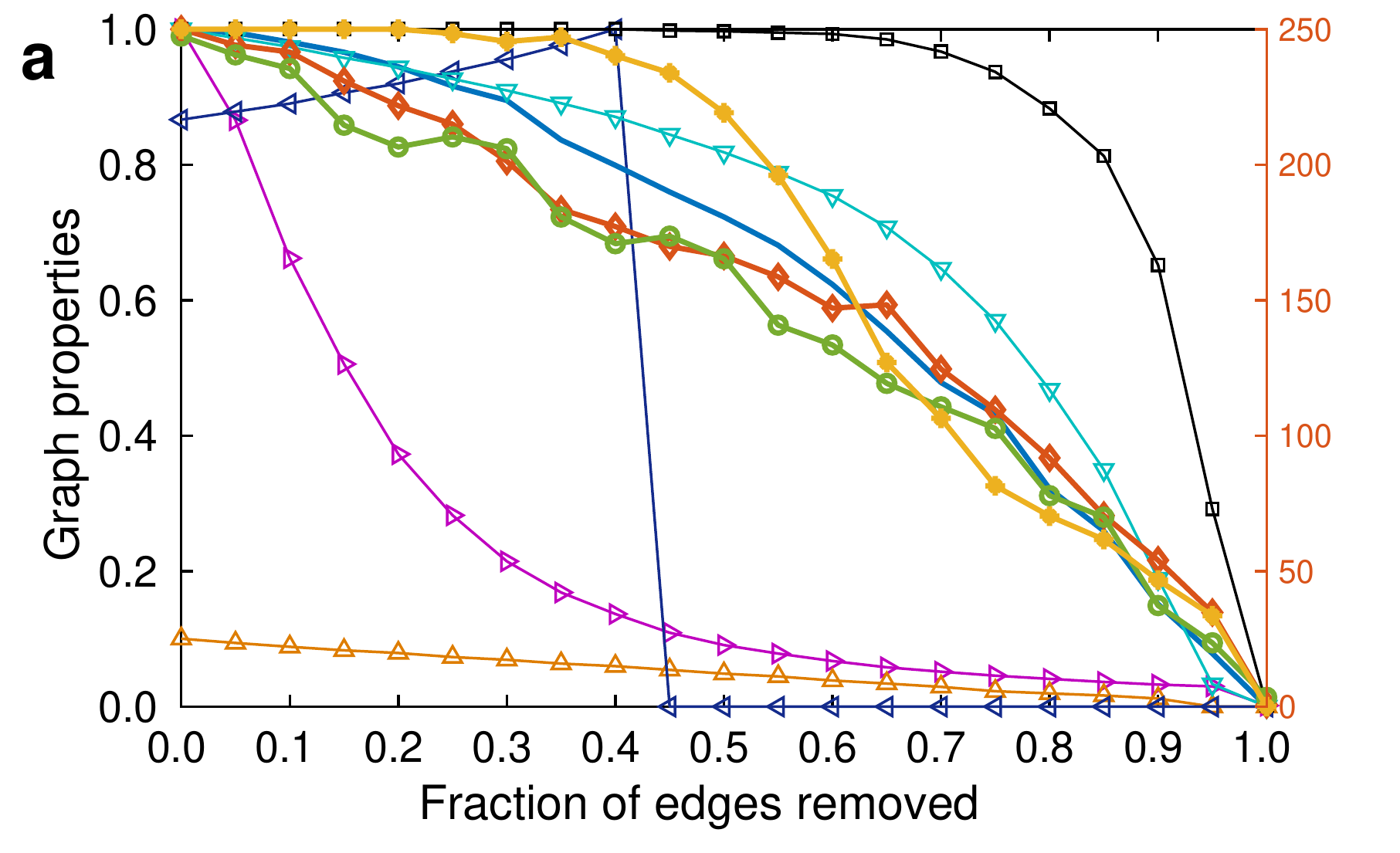}}
{\includegraphics[width=0.48\linewidth]{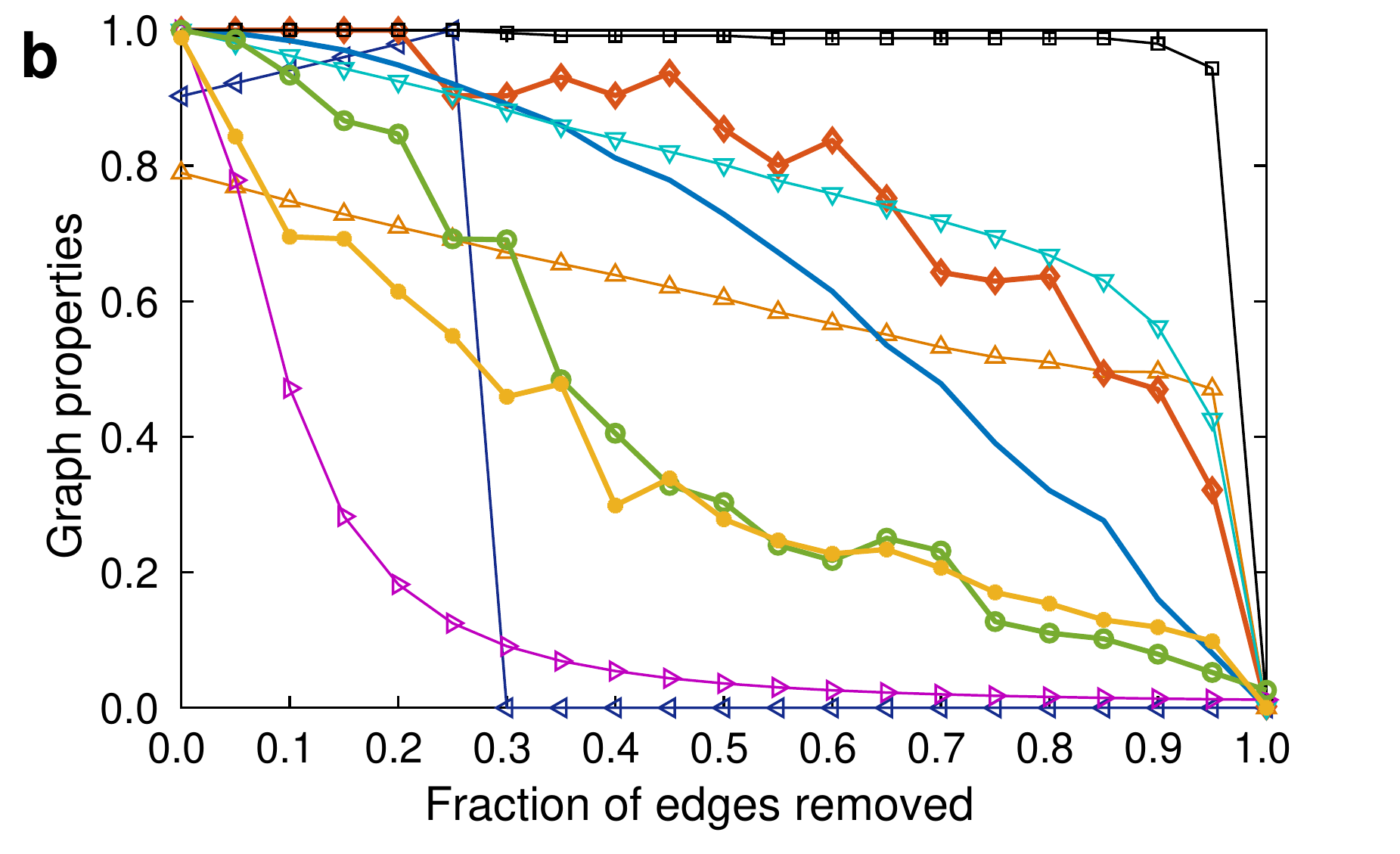}}
\caption{Variation of MASS and other variables under weight thresholding, for a synthetic network generated by the LFR benchmark (a) and the world trade network (b) with randomized weight and $0$ weight-degree correlation . The "$|$maxComponent$|$" curve represents the fraction of vertices remaining in the largest connected component.}
\label{nocorr}
\end{figure*}

\added{To demonstrate the effect of weight thresholding on networks with no weight-degree correlation, we rerun the experiment on the synthetic and world trade network with randomized edge weight. In both cases, MASS and mesoscopic structures fall quickly as the edges are removed (Fig.~\ref{nocorr}).}

\section{Theoretical properties of MASS}
In the paper \cite{Koutra_DeltaCon_2016}, the authors proposed several theoretical axioms for graph similarity measures. Here, we adapt them to the specialized task of comparing $G$ with its thresholded subgraph $\tilde{G}$. We interpret the axioms as desired properties that good subgraph similarity metrics (SSM) must satisfy. We first introduce the full list of axioms with interpretations.
\begin{enumerate}
 \item Zero-Identity: The SSM returns $0$ if $\tilde{G}$ is an empty graph; $1$ if $\tilde{G}$ is identical to the original graph.
 \item Monotonicity: The SSM (non-strictly) monotonically decreases or increases as we threshold out more edges.
 \item Robustness: The SSM will not drop to zero from relative large values by removing a single edge, even if the graph becomes disconnected (unless it becomes an empty graph).
 \item Submodularity: Removing the same set of edges has a greater impact on the subgraph similarity measure for smaller graphs.
 \item Weight awareness: When thresholding out a single edge, the greater the edge weight, the greater the impact on the subgraph similarity measure.
 \item Structure awareness: The SSM suffers from a greater impact if thresholding creates disconnected components.
\end{enumerate}

Next, we demonstrate that the proposed MASS measure satisfies these axioms. Recall that we define MASS as
$$\sigma_{min}(\tilde{G}, G)  = 1 - \dfrac{\lambda_N^{\varDelta}}{\lambda_N}\;,$$
where $\lambda_N^{\varDelta}$ is the largest eigenvalue of the Laplacian of the difference graph $\varDelta G = \{V, \varDelta E = E - \tilde{E}, \varDelta W = W -\tilde{W}\}$. 

The first axiom requires that the subgraph similarity measure returns $0$ for a completely thresholded graph and $1$ for the original graph. 
\begin{property}[Zero-Identity]
$\sigma_{min} (\varnothing, G) = 0$ and $\sigma_{min} (G, G) = 1$, where $\varnothing$ represents an empty graph $\varnothing = \{V, \varnothing, 0_{N,N}\}$. 
\end{property}
\begin{proof}
The Zero property is trivially satisfied as $\varDelta G = G$ for $\tilde{G} = \varnothing$ and thus $\lambda_N^{\varDelta} = \lambda_N$. The Identity property holds because $\varDelta G = \varnothing$ for $\tilde{G} = G$. According to spectral graph theory, all eigenvalues of the Laplacian matrix are $0$ for an empty graph with $N$ connected components and we have $\lambda_N^{\varDelta} = 0$.
\end{proof}

The second axiom requires the subgraph similarity measure to be monotonically decreasing as we threshold out more and more edges.
\begin{property}[Monotonicity]
$\sigma_{min} (A, G) \le \sigma_{min} (B, G)$ if $A$ is a subgraph of $B$, where $A,B$ are both subgraphs of the original graph $G$.
\end{property}
\begin{proof}
By complement, we know that $\varDelta B$ is a subgraph of $\varDelta A$. Assume that $\lambda_N^{\varDelta B} = v^T \varDelta L_B v$, where $v$ is the corresponding unit length eigenvector. Because of the monotonicity of the Laplacian quadratic form, we have $v^T \varDelta L_A v \ge \lambda_N^{\varDelta B}$. We also have $\lambda_N^{\varDelta A} \ge v^T \varDelta L_A v \ge \lambda_N^{\varDelta B}$. Therefore $\sigma_{min} (A, G) \le \sigma_{min} (B, G)$.
\end{proof}

Monotonicity alone does not prevent degeneracy when the subgraph becomes disconnected. The third axiom thus states that the subgraph similarity measure will not drop to zero from relative large values by removing an arbitrary edge. A general proof for weighted graphs is difficult to formulate. Here we focus on simple graphs. 
\begin{property}[Robustness]
Let $A$ be a subgraph of $B$ by removing an arbitrary edge, where $A,B$ are both subgraphs of the original graph $G$, and all graphs are unweighted. If $\sigma_{min} (B, G) > 0.5$, we have $\sigma_{min} (A, G) >0$.
\end{property}
\begin{proof}
Since $\sigma_{min} (B, G) =  1 - \dfrac{\lambda_N^{\varDelta B}}{\lambda_N} > 0.5$, we have $2 \lambda_N^{\varDelta B}<\lambda_N$. The largest eigenvalue of the Laplacian of $\varDelta B$ is bounded on both sides by
$$\max_i d_i^{\varDelta B} +1 \le\lambda_N^{\varDelta B} \le \max_{i,j\in \varDelta B} (d_i^{\varDelta B} + d_j^{\varDelta B})\;,
$$
where $d_i^{\varDelta B}$ denotes the degree of vertex $i$ in $\varDelta B$. Without loss of generality, we assume that $\varDelta A$ is $\varDelta B$ plus the edge $(u,v)$. If $(u,v)$ becomes the new maximizer for the upper bound of the Laplacian of $\varDelta A$, we have
$$\lambda_N^{\varDelta A} \le d_u^{\varDelta A} + d_v^{\varDelta A} = d_u^{\varDelta B} + d_v^{\varDelta B} + 2 \le 2 \lambda_N^{\varDelta B}<\lambda_N \;. 
$$
If the maximizer is $(u',v') \neq (u,v)$, we instead get
$$\lambda_N^{\varDelta A} \le d_{u'}^{\varDelta A} + d_{v'}^{\varDelta A} \le d_{u'}^{\varDelta B} + d_{v'}^{\varDelta B} + 1 < 2 \lambda_N^{\varDelta B} <\lambda_N\;. 
$$
Therefore, we always have 
$$\sigma_{min} (A, G) = 1 - \dfrac{\lambda_N^{\varDelta A}}{\lambda_N} > 0
$$
\end{proof}
Monotonicity and smoothness concerns different thresholding on the same graph $G$. We can similarly derive submodularity for the same thresholding on different graphs. In other words, removing the same set of edges has a greater impact on the similarity measure for smaller graphs.
\begin{property}[Submodularity]
Let $A$ be a subgraph of $B$. For any common thresholding on both graphs such that $\varDelta A = \varDelta B \in A$, we have $\sigma_{min} (\tilde{A}, A) \le \sigma_{min} (\tilde{B}, B)$.
\end{property}
\begin{proof}
Assume $\lambda_N^{A} = v^T L_{A} v$, where $v$ is the corresponding unit length eigenvector. Because of the monotonicity of the Laplacian quadratic form, we have $v^T L_{B} v \ge \lambda_N^{A}$. We also have $\lambda_N^{B} \ge v^T L_{B} v \ge \lambda_N^{A}$. Since $\varDelta A = \varDelta B$, we get $\lambda_N^{\varDelta A} = \lambda_N^{\varDelta B}$. Therefore,
$$\sigma_{min} (\tilde{A}, A) = 1 - \dfrac{\lambda_N^{\varDelta A}}{\lambda_N^A} \le 1 - \dfrac{\lambda_N^{\varDelta B}}{\lambda_N^B} =\sigma_{min} (\tilde{B}, B).
$$
\end{proof}

The fifth axiom asserts that when thresholding out a single edge, the greater the edge weight, the greater the impact on the similarity measure.
\begin{property}[Weight Awareness]
Let $A,B$ be different subgraphs of the original graph $G$ by removing a single edge, with the edge $(u,v)\notin A$ (but $\in B$), and $(u',v')\notin B$ (but $\in A$). If the edge weights follow $W_{uv} \ge W_{u'v'}$, we have $\sigma_{min} (A, G) \le \sigma_{min} (B, G)$.
\end{property}
\begin{proof}
By complement, we know that $\varDelta A$ and $\varDelta B$ both consist of a single edge and $W_{uv}^{\varDelta A} \ge W_{u'v'}^{\varDelta B}$. The largest eigenvalue of the Laplacian of a single edge graph is simply  $\lambda_N^{\varDelta A} = 2 W_{uv} \ge 2 W_{u'v'} = \lambda_N^{\varDelta B}$, and we thus have 
$\sigma_{min} (A, G) = 1 - \dfrac{2 W_{uv}}{\lambda_N}  \le 1 - \dfrac{2 W_{u'v'}}{\lambda_N} = \sigma_{min} (B, G)$. 
\end{proof}

The last property provides an important structural constraint. If thresholding creates disconnected components, it should have a greater impact on the similarity measure. Axiomatizing this property in its most general form is difficult, we thus focus on an intuitive special case: the unweighted Barbell graph.

\begin{property}[Structure Awareness]
Let graph $G$ be an unweighted graph with two non-overlapping stars of equal size $N/2$ connected by a single edge $(u,v)$, where $u$ and $v$ are the two center vertices. Assume that $A,B$ are subgraphs of $G$, which differs only by swapping a single edge, with the edge $(u,v)\notin A$ (but $\in B$), and $(u',v')\notin B$ (but $\in A$). Then we have $\sigma_{min} (A, G) \le \sigma_{min} (B, G)$.
\end{property}
\begin{proof}
By complement, we know that $\varDelta A$ is composed of two connected stars while $\varDelta B$ consists of two disconnected stars. Without loss of generality, let us assume that $|\varDelta A| = |\varDelta B| = k$, and that the two stars in $\varDelta B$ have sizes $k_1 \le k_2$, with $k=k_1+k_2$. Since the largest eigenvalue of Laplacian of an unweighted $k$-star is exactly $k$, and the spectrum of the Laplacian of a disconnected graph is simply the union of those of its components, we have $\lambda_N^{\varDelta B} = k_2$.

Without loss of generality, assume that the bridge edge $(u,v)$ in $\varDelta A$ has its endpoint $u$ in the bigger component, which is of size $k_2$. Vertex $u$ therefore has a degree that is at least $k_2$ (It will equal $k_2 +1$ if $(u',v')$ is in the opposite star component). Vertex $u$ and its neighbors thus form a $(k_2+1)$-star subgraph of $\varDelta A$. Since the largest eigenvalue of Laplacian of an unweighted $(k_2+1)$-star is exactly $k_2+1$, by the monotonicity of the Laplacian quadratic form, we have $\lambda_N^{\varDelta A} \ge k_2+1 \ge \lambda_N^{\varDelta B}$. Therefore, $\sigma_{min} (A, G) \le \sigma_{min} (B, G)$.
\end{proof}
\end{document}